\documentclass[twocolumn]{aastex631}
\pdfoutput=1 
\usepackage{amsmath,amstext,amssymb,mathtools,graphicx,color}
\usepackage{booktabs, multirow} 

\usepackage{graphicx}	
\usepackage{amsmath}	
\usepackage{amssymb}	
\usepackage{xcolor}
\usepackage{lipsum}
\usepackage[normalem]{ulem}
\usepackage{graphicx}
\usepackage[caption=false]{subfig}

\submitjournal{ApJ}

\shorttitle{Energy Sources of Ic-BL SNe}
\shortauthors{Niblett et al.}

\graphicspath{{./}{figuresold/}}

\begin{document}

\title{Studying the Power Sources Behind Type Ic Supernovae}

\correspondingauthor{Annabelle Niblett}
\email{annabelle.niblett@gmail.com}

\author[0009-0005-2363-9274]{Annabelle E. Niblett}

\affiliation{Department of Physics \& Astronomy, Dartmouth College, Hanover, NH 03755 USA}
\affiliation{Center for Nonlinear Studies, Los Alamos National Laboratory, Los Alamos, NM 87545 USA}

\author[0009-0009-3248-5253]{Daniel A. Fryer}
\affiliation{Center for Nonlinear Studies, Los Alamos National Laboratory, Los Alamos, NM 87545 USA}

\author[0000-0003-2624-0056]{Christopher L. Fryer}
\affiliation{Center for Nonlinear Studies, Los Alamos National Laboratory, Los Alamos, NM 87545 USA}

\begin{abstract}
Astrophysical transients can be powered by a broad range of energy sources including shock-heating (internal and external shocks), decay of radioactive isotopes, and long-lived central engines (magnetar and fallback). The dominant energy source for astrophysical transients depends on the nature of the explosive engine and its progenitor. To model all transients, light-curve codes must include all of these energy sources. Here we present a supernova light-curve code implementing analytic source models to compare the role of different energy sources in these transients. To demonstrate the utility of this code, we conduct an extensive study of type Ic broad-line supernovae. A diverse set of energy sources have been linked to Ic broad-line supernovae making them an excellent candidate for this light-curve code. In this paper, we explore which features of the explosion (mass, velocity, etc.) affect the type Ic supernovae light-curves, focusing on shock-interaction and radioactive decay energy sources. Although the explosion properties under both energy sources can be tuned to match the peak emission, matching the light-curve evolution in many Ic broad-line supernovae requires fine-tuned conditions. We find that shock interactions in the stellar wind are likely to be the dominant energy source at peak for these supernovae.
\end{abstract}


\section{Introduction} 

The energy sources powering supernovae can arise from a broad range of physical conditions: internal shocks, external shocks (interactions with clumpy wind media, shells from mass transfer events, etc.), energy deposition from the decay of radioactive isotopes, and continued emission from the collapsed core (magnetar or fallback accretion). Internal shocks occur as the supernova blast wave propagates through the star.  These shocks can be much more complex than the forward (and reverse) shocks produced in spherical explosions. Convective motion in the star leads to density perturbations that can drive asymmetries in the already asymmetric supernova blast wave~\citep{2022ApJ...933..164G}. Interactions between the blast wave and the circumstellar medium can drive heating at later times, powering extended shock-heated emission. These shocks heat the ejecta, powering the observed transient emission.  The internal shock energy source is often believed to be the power source behind type II supernovae and a subset of superluminous supernovae~\citep{2013ApJS..204...16F,2017ApJ...850..133D,2017hsn..book..403S,2019ApJ...876..148C}. 

The decay of radioactive isotopes is another common energy source, especially in compact explosions, where adiabatic expansion minimizes the role of internal shocks. Nearly half of the ejecta in a type Ia supernova is radioactive $^{56}$Ni~\citep{2013MNRAS.429.1156S} and the decay of $^{56}$Ni and its daughter product $^{56}$Co releases energy (in the form of gamma rays and positrons) that heats the ejecta, powering the observed supernovae. The Arnett law~\citep{1982ApJ...253..785A}, which captures this energy deposition in a simple analytic model, has been used to infer the $^{56}$Ni content of these supernovae from their peak emission.

Long-term energy sourcing from the central engine can also contribute to the observed supernova emission. Magnetic fields developing in the newly-formed neutron star can extract the neutron star’s rotational energy, augmenting the emission from supernovae. Fast-rotating magnetars are believed to power a fraction of superluminous supernovae~\citep{2010ApJ...717..245K}. Accretion onto the neutron star, e.g. supernova fallback~\citep{2009ApJ...699..409F}, can also augment the supernova emission~\citep{2013ApJ...772...30D,2019ApJ...880...21M}. Both of these sources deposit their energy in the innermost ejecta and, as such, will likely only contribute to the late time emission from a transient unless the transient is low-mass. 

Internal shocks power supernovae from extended red giant or supergiant stars (type II core-collapse supernovae). Radioactive decay powers supernovae from compact white dwarf stars (type Ia thermonuclear supernovae). But what about type Ib/c supernovae? These supernovae arise from relatively compact helium or carbon/oxygen Wolf-Rayet stars. As such, these stars will adiabatically cool as their ejecta expand, limiting the peak flux powered by internal shocks. Radioactive power also has its issues explaining Ib/c supernovae. The energy released from the decay of $^{56}$Ni produced in the innermost ejecta of these explosions will take time to reach the supernova photosphere and is likely to drive peak emission well after the observed 10-20\,d peak timescales. 

One way to explain the rise time of these supernovae is to mix out the $^{56}$Ni into the outer layers of the ejecta through a highly asymmetric explosion. Coupled with fast velocity speeds and low ejecta masses, this mixing will allow the $^{56}$Ni to power the light-curves at early times. Observations of the light-curve before and at peak is an excellent constraint on both the total $^{56}$Ni production and the amount of mixing. 

Although internal shocks can't explain the light-curves of Ib/c supernovae, the emission from these supernovae can be powered by shock interactions of the blast wave with the circumstellar medium: e.g. shells from binary mass ejection or clumpy winds. The progenitors of type Ib/c supernovae experience extensive mass loss episodes. For many of these progenitors, binary mass ejection is responsible for the removal of the hydrogen (and perhaps helium) envelopes, often ejected in shells~\citep{1992ApJ...391..246P,2013A&ARv..21...59I}. These stars also drive strong, inhomogeneous winds. The supernova blast wave will produce shocks as it propagates through this wind~\citep[for a review, see][]{Fryer_2020}. For an increasing number of observed supernovae, strong evidence is seen of shock interactions that drive late-time emission~\citep{2017hsn..book..403S,2024ApJ...973...14C}. But if these shocks occur at early times, they can dominate the supernova emission at peak brightness. 

In this paper, we study the role of different energy sources on supernovae light-curves focusing on the progenitors of Ic supernovae. We focus on the highly energetic Ic broad-line (Ic-BL) supernovae (also known as SNe). In this paper, we first describe our computational methods and its implementation of the different energy sources in Section~\ref{sec:code}. Our base Ic-BL progenitor is described in Section~\ref{sec:grid} along with the variations in this base model that make our grid of simulations. Section~\ref{sec:results} presents the results of this grid of simulations, describing the processes driving the trends in the early-time light-curves. We conclude with a discussion of the implication of these results on our analysis of Ic-BL supernovae.

\section{Computational Tools}
\label{sec:code}

Here we describe both the open-source Supernova light-curve code {\it SNLC} and a suite of post-processing tools to produce band and spectral information from these explosions. This section describes the transport scheme as well as the energy sources for the {\it SNLC} followed by a brief discussion of the post-process tools\footnote{At this time, access to {\it SNLC} and subsequent tools is by request to fryer@lanl.gov.}.

\subsection{Transport}
\label{sec:transport}

For radiation transport, we use a single-energy group, flux-limited diffusion transport code. We have implemented a number of flux limiters but, for these calculations, we use the simple flux limiter discussed in \cite{1994ApJ...435..339H} that limits the radiation diffusion to the free-streaming limit capturing both the free-streaming and diffusive limits in the calculation. We assume the radiation and matter are in equilibrium and their distributions are described by a single temperature (the ``1-T'' limit used in many stellar evolution codes).  This is among the simplest methods for radiation transport.  A wide range of radiation hydrodyanamics techniques are used in supernovae~\citep{2023arXiv231216677F}.  Flux-limited diffusion is a type of moment closure methods and these have been used by a number of radiation-hydrodynamic supernova light-curve codes~\citep{2023arXiv231216677F}.  For example, \cite{2013ApJS..204...16F} use a multi-group flux-limited diffusion in their radiation-hydrodynamics calculations (our model is single-group).   The strength of our method is that it is extremely fast, ideally-suited for studies focusing on understanding trends in the physics.

The method used for radiation transport ranges widely between different astrophysics transport codes~\citep[e.g.][]{2022A&A...668A.163B} with different levels of approximation.  Flux-limited diffusion is one of the simplest moment closure techniques used in the literature.  Moment closure techniques include $M_N$ and variable Eddington factor/tensor techniques.  Roughly, the difference between these schemes relies on the number of moments used before closure.  For example, \cite{2011ascl.soft08013B} uses a higher-order variable Eddington factor technique which captures more details of the angular distribution of the radiation, again using multi-group.   Other techniques discretize the angle using spherical harmonics or discrete ordinate methods (often called ``Boltzmann'' transport in the astrophysical community).  Finally, Monte-Carlo schemes are typically considered the gold standard for full transport.  All of these techniques have their strengths and weaknesses(e.g. Monte-Carlo suffers from low-packet noise), but if we compare the methods used in \citep{2022A&A...668A.163B}, the difference between different methods can be less than the differences caused by schemes to implement the opacities.  For a review of transport techniques, we point the reader to \citep{2004rahy.book.....C,2023arXiv231216677F}. 

Opacities are implemented in the {\it SNLC} code using a linear interpolation (in both density and temperature) of Rosseland mean opacity tables from the LANL opacity database~\citep{2016ApJ...817..116C} (a ``line-binned'' approach). For our Ic supernova, we assume the composition is dominated by carbon and oxygen (50/50 C and O).  Binned, equilibrium opacities are used by a number of groups including SuperNu~\citep{2013ApJS..209...36W}. The binned approach has been compared to other line sampling and expansion opacity methods~\citep{2014ApJS..214...28W,2020MNRAS.493.4143F,2022A&A...668A.163B}.  However, some groups have argued that expansion opacities can be a factor of 2-5 higher for C/O stars~\citep{2017hsn..book..843B}.  Using single-group (gray), instead of multi-group, opacities can also have a similar effect.  As such, we will include a few simulations that artificially raise the opacity in our calculations to determine its effect on our results.

These opacities assume local thermodynamic equilibrium, consistent with our presumption that the radiation and matter is tightly coupled. Although local thermodynamic equilibrium is never truly correct, 
the high densities at the early times studied here typically lead to solutions that are well approximated by these equilibrium assumptions.  In addition, the rapid motion of the photosphere means that errors in the opacity have a reduced effect on the light-curve evolution.

\subsection{Energy Deposition Terms}

The {\it SNLC} includes prescriptions for 4 different energy sources:  radioactive decay (for supernovae, the typical radioactive isotope $^{56}$Ni), external shock-heating, magnetar/pulsar power, and fallback accretion. This section discusses the numerical implementation of each of these terms.

\subsubsection{Ni decay}

Radioactive decay provides a partial power source for a broad range of astrophysics from kilonovae to supernovae. For supernovae (whether powered by the convective or accretion disk engine), $^{56}$Ni is produced in abundance and it is the dominant radioactive power source. $^{56}$Ni decays to $^{56}$Co that then decays to $^{56}$Fe. The change in mass over a given timestep ($dt$) is:
\begin{eqnarray}
 dM_{^{56}Ni} = -M_{^{56}Ni} (1-e^{-dt/\tau_{^{56}Ni}}) \\
 dM_{^{56}Co} = -dM_{^{56}Ni} - M_{^{56}Co} (1-e^{-dt/\tau_{^{56}co}}) \\
 dM_{^{56}Fe} = -dM_{^{56}Co}
\end{eqnarray}
where $\tau_{^{56}Ni,Co}$ are the half lives of $^{56}$Ni and $^{56}$Co respectively. For the decay timescales, we use values from the National Nuclear Data Center, information extracted from the NuDat database, \url{https://www.nndc.bnl.gov/nudat/}. 

We also use the Q-values from this database to calculate the energy released in the decay. This energy is assumed to be deposited in situ. It must then diffuse out to the photosphere. As the ejecta evolves, the gamma rays and positrons may escape without depositing their energy. We estimate this escape fraction by calculating the optical depth of the gamma rays ($\tau_\gamma$) at each zone and reduce the energy deposited by gamma rays through this escape:
\begin{equation}
 E_{\rm dep} = (E_{Ni}+E_{Co}) (1-e^{-\tau_{\gamma}})
 \label{eq-dep}
\end{equation}
where $E_{^{56}Ni}$ and $E_{^{56}Co}$ are the energies released by $^{56}$Ni and $^{56}$Co decay. Here we assume that either the energy is deposited in situ or it escapes. We don't do a full transport of either the gamma rays or the positrons. In our simplest model, we assume the positrons escape with the gamma rays. But we also include a prescription that assumes, for the duration of our simulations, that the positrons are trapped, always depositing the positron energy in situ:
\begin{equation}
 E_{\rm dep} = (E^\beta_{Ni}+E^\beta_{Co}) + (E^\gamma_{Ni}+E^\gamma_{Co}) (1-e^{\tau_{\gamma}})
 \label{eq-dep2}
\end{equation}
where $E^\beta$ is the energy released in positrons and $E^\gamma$ is the energy released in gamma rays.  

Our prescription is simplistic:  either the gamma-rays/positrons are deposited in situ or they escape.  Although most groups make this assumption for positrons, many light-curve calculations transport the gamma-rays.  As the region becomes less dense, transport models allow the gamma-rays to deposit their energy across a distributed region.  Electron/positron transport is more difficult (requiring the understanding of interactions with magnetic fields).  To our knowledge, no group includes a distributed energy distribution from positrons.  For our implementation of this energy deposition, we under-estimate the energy deposited.  For instance, gamma-rays moving inward may deposit much more energy as they scatter before they escape (if they escape at all).  By under-estimating the energy deposition, our light-curves will decay more rapidly than light-curves with more complete models.  As we shall see in section~\ref{sec:comparison}, our Ni-only powered light-curves suffer from too-long decay times even for our models.  This issue will be more extreme with more accurate deposition models.

$^{56}$Ni is primarily produced in the innermost ejecta. If it is not mixed out in the ejecta, the amount of slow moving ejecta can drastically alter the role its decay plays in powering supernovae. If the $^{56}$Ni is slow moving, the energy released from its decay can power the supernova out to late times. Measuring the amount of slow-moving $^{56}$Ni will constrain the extent of late-time power from these supernovae. The distribution of the $^{56}$Ni can be estimated from the Doppler broadening of the $^{56}$Fe lines if we can differentiate the iron from $^{56}$Ni decay and that from the progenitor star at formation (less important at the lower metallicities of Ic-BL supernovae).

\subsubsection{Shock Interactions}

Most of the energy in a core-collapse supernova is in the form of kinetic energy. When a shock decelerates, this energy is converted into thermal energy which can power supernovae. For type II supernovae, the light-curve through peak is entirely powered by the shock deceleration that occurs as the shock propagates through the hydrogen envelope. As the front decelerates, material from the explosion can pile up, causing a reverse shock to heat a considerable portion of the ejecta. This internal shock-heating can be implemented simply by raising the temperature at the shock~\citep{2017ApJ...850..133D}. The {\it SNLC} code includes a temperature factor that can increase the shock-heating to mimic the effects of internal shock-heating, particularly from the complex shock structures produced when we include both asymmetric explosions and asphericities in the convective hydrogen envelope.

A number of supernova transients exhibit evidence of strong external shock interactions from a subset of type II supernovae~\citep[for a review, see]{2017hsn..book..403S} and superluminous supernovae~\citep[e.g.][]{2019ApJ...874...68C} to peculiar supernovae, e.g. with bright ultraviolet or X-ray emission~\citep[e.g.][]{2023Univ....9..218B,2024PASA...41...59P}. This shock-heating can occur at any time after the blast wave breaks out of the star and is the result of a wide range of interactions: clumpy stellar wind, binary shell ejecta, and binary companions.  Although detailed theoretical studies argued that these outflows should have complex structures, there is a growing evidence that at least some sort of asymmetry must exist in the ejecta to explain existing light curves~\citep[e.g.][]{2025arXiv250113619R,2025arXiv250113621R}.  To implement this shock-heating, we include formulae to decelerate the shock and consider the effects of the reverse shock. We have two basic models for this deceleration mimicking strong shell and clumpy wind/porous shell interactions.

For the strong shell interactions, we assume a pronounced deceleration of the shock based on a Sedov blast wave solution where the velocity of the forward shock decelerates dramatically with time. The deceleration depends both on the density gradient and mass in the shell. In a smooth density profile, the velocity evolution depends on the density gradient and is a power law based on the Taylor–von Neumann–Sedov similarity solution~\citep{1959sdmm.book.....S}:
\begin{equation}
 v_{\rm shock} \propto v_0 (t/t_0)^{(\alpha-3)/(5-\alpha)}
\end{equation}
where $\alpha$ denotes the density gradient ($\rho \propto r^{-\alpha}$), $v_{0}$ is the initial velocity when the shock starts decelerating, and $t_0$ is the time at the onset of the deceleration. Depending upon the density profile of the shell or wind, the value for $\alpha$ can range anywhere from 0 (flat density shell) to 2 (wind profile: $\rho \propto \dot{M}_{\rm wind}/(4 \pi r^2 v_{\rm wind})$) yielding a velocity evolution from $t^{-3/5}$ to $t^{-1/3}$. If the shell is porous or the wind is clumpy, the velocity deceleration will be much less extreme. We implement this porous shell/clumpy wind external shock-heating through two different models: a sharp boundary condition and a smoothed boundary condition.

For the sharp boundary condition, we use a simple linear velocity reduction:
\begin{equation}
 v_{\rm shock} = v_{\rm orig} [1 - \beta (t-t_{\rm begin})/(t_{\rm end}-t_{\rm begin})]
 \label{eq:vshock}
\end{equation}
where $v_{\rm orig}$ is the original velocity at the front of the supernova blast wave and $t_{\rm begin},\ t_{\rm end}$ mark the beginning and end times of the deceleration (e.g. entrance and exit of a dense, clumpy wind or shell). $\beta$ is usually set to a small value corresponding to the final deceleration of the forward shock. As we shall see, for these energetic explosions, a few \% deceleration of the shock velocity can power a luminous supernova light-curve.

As the forward shock decelerates, material begins to pile up against this front, causing it to decelerate as well and produce a reverse shock. We include this reverse shock by limiting the velocity of any inner ejecta to equal or less than that of the forward shock:
\begin{equation}
 v_{k-1} = min(f_{\rm reverse} v_{k}, v_{k-1})
\end{equation}
where $v_k$ is the velocity of zone $k$ and $f_{\rm reverse}$ is some fixed factor (usually between 0.999-1.0). As the forward shock slows down, so does the material immediately behind it. The kinetic energy released in this deceleration is immediately deposited in the decelerating zone's thermal energy.

Equation~\ref{eq:vshock} produces a strong spike in the luminosity due to the sudden deceleration. But a more porous shell, clumpy wind, or other ejecta asymmetry can cause a more gradual deceleration and hence shock-heating. Although our 1-dimensional calculations can not capture the effects of large-scale asymmetries, we can capture the stochasticity in the circumstellar medium caused by these asymmetries. For these shocks, we implemented a sigmoid solution:
\begin{equation}
 v = v_0 \left(\frac{\delta v}{1+e^{-k(t-t_{\rm begin})}} + (1-\delta v) \right)
 \label{eq-sig}
\end{equation}
We solved for $t_{begin}$ and $k$ analytically in Wolfram Mathematica, setting the total velocity deceleration in the shock to produce different peak luminosities. We can then adjust the shock-heating duration, onset time, and steepness to produce a range of light-curves. These variations would be caused by the conditions in the circumstellar material. Without a complete understanding of this environment, we are free to choose any values for these parameters, allowing us adjust parameters to achieve best fits to existing Ic-BL light-curves.  

\subsubsection{Magnetar or Pulsar Energy Injection}

Neutron stars formed in stellar collapse can develop strong magnetic fields and, if born rapidly rotating, these magnetic fields can extract the rotational energy to power supernovae explosions. How important this energy injection is to the supernova light-curve also depends on the extent of the magnetic field burial, and its subsequent emergence time~\citep{2004MNRAS.351..569P,2019EPJA...55..132F}. Our code contains the option of an energy source from a magnetar formed some set emergence time after the launch of the explosion. Our prescription assumes a neutron star with a dipole magnetic field yielding an energy injection rate of~\citep{2014MNRAS.439.3916M}:
\begin{equation}
 L_{\rm mag} \simeq 6\times10^{49}B^{2}_{15}P^{-4}({\rm ms})(1+t/t_{\rm mag})^{-2}\mathrm{erg\:s^{-1}}
\end{equation}
where $B_{15}$ is the magnetic field in $10^{15}$G, $P$ is the spin period of the magnetar in ms, and $t_{\rm mag}$ is the spin-down timescale:
\begin{equation}
t_{mag} \simeq 0.14B^{-2}_{15}P^{2}({\rm ms})\mathrm{h}.
\end{equation}
The magnetar energy will be deposited in the innermost supernova ejecta by adding both momentum and energy. In our current model, we only include the energy injection and assume this energy is immediately deposited into the innermost zone of the supernova ejecta. However, we also include the option to delay the energy injection assuming that there is a delay in the emergence of the magnetic field.  In scenarios where the neutron star collapses to a black hole after a set time, the magnetar power source will abruptly stop.  Such an option is also implemented into the code.

In this paper, our focus is on the peak luminosity of Ic-BL supernovae. As we shall see, even for our $^{56}$Ni-powered runs, it is difficult to explain the peak luminosity unless the $^{56}$Ni is mixed into the ejecta. Central power sources like a magnetar are unlikely to affect the peak luminosity for our models and hence this power source will not be the focus of this study. But, especially at late times, this power source can be important.

\subsubsection{Fallback}

The supernova shock plows through the star and accelerates it, causing the ejection of the star and this ejecta forms the supernova blast wave. But not all of the outward moving material is ejected. As the supernova shock pushes outward, some material decelerates below the escape velocity and collapses back onto the compact object formed in the collapse~\citep{2009ApJ...699..409F}. When this material accretes onto the compact remnant, a fraction of the gravitational potential energy released in the accretion can power an additional energy source, deposited into the innermost ejecta. For this energy source, we use the sourcing term from \cite{2019ApJ...880...22W}:
\begin{equation}
 L_{\rm fallback} = \eta G M_{\rm CR}/r_{\rm CR} \dot{m}^0_{\rm fallback} (t/t_0)^{-5/3}
\end{equation}
where $G$ is the gravitational constant, $M_{\rm CR}$ and $r_{\rm CR}$ are the compact mass and radius respectively, $\eta$ is the efficiency at which the potential energy released powers the light-curve, and $\dot{m}^0$ and $t_0$ are the initial fallback accretion rate starting at time $t_0$ that then decreases with time ($t$).   The $-5/3$ power-law decay with time for the accretion rate derives from analytic estimates of neutron star accretion~\citep{1989ApJ...346..847C} roughly matching a broad range of simulations studies.

As with the pulsar/magnetar source, although this fallback accretion will inject both energy and momentum to the system, we currently only include the energy injection. This energy is deposited in the innermost zone of the ejecta and is unlikely to contribute to the emission at peak. 

\subsection{Post-Process}

Our transport code is currently gray (single energy). We use post-process codes to produce spectra and band emission from these calculations. We have two different post-process codes: 1) {\it lphot} calculates the photosphere and band emission assuming a blackbody emission using only the conditions at the position of a Rosseland mean opacity photosphere; and 2) {\it lprof} uses the entire ejecta profile to calculate the the emission. {\it lphot} uses the density and temperature at the photosphere to calculate the opacity and Kirchhoff's law of thermal radiation assuming detailed balance to calculate the opacity-mediated blackbody emission. Here, we implement a simple photospheric velocity to incorporate Doppler effects. {\it lprof} calculates emission and absorption at every zone in the stellar ejecta and can be used to calculate detailed spectra with the only caveat that the temperature/density conditions used to calculate the spectra are based on a gray flux-limited diffusion transport solution.

\section{Simulation Grid}
\label{sec:grid}

For our base model, we use the model ``M25aE7.42'' from \cite{2018ApJ...856...63F}. In this paper, the supernova explosions were driven using a prescription to mimic the convective engine in 1-dimension. This is an impulse explosion (energy deposition in less than 1\,s). This explosion will be on par with a short-lived accretion-disk wind engine, but a long-lived disk model will likely have a different profile. This model is a strong explosion of a 25\,M$_\odot$ progenitor, with an explosion energy when the shock propagates through its hydrogen envelope of $7.42 \times 10^{51} {\rm \, erg}$. In this paper, we remove the hydrogen and helium envelopes, studying the explosion when it breaks out of the Carbon/Oxygen (C/O) core. Without energy losses due to deceleration (causing a slight amount of fallback) in the shock propagating through these envelopes, the corresponding base explosion energy at this breakout time is $7.45 \times 10^{51} {\rm \, erg}$. We use this explosion model to define the base density, entropy, temperature and velocity profiles of our ejecta.

\begin{figure}
 \centering
 \includegraphics[width=\linewidth]{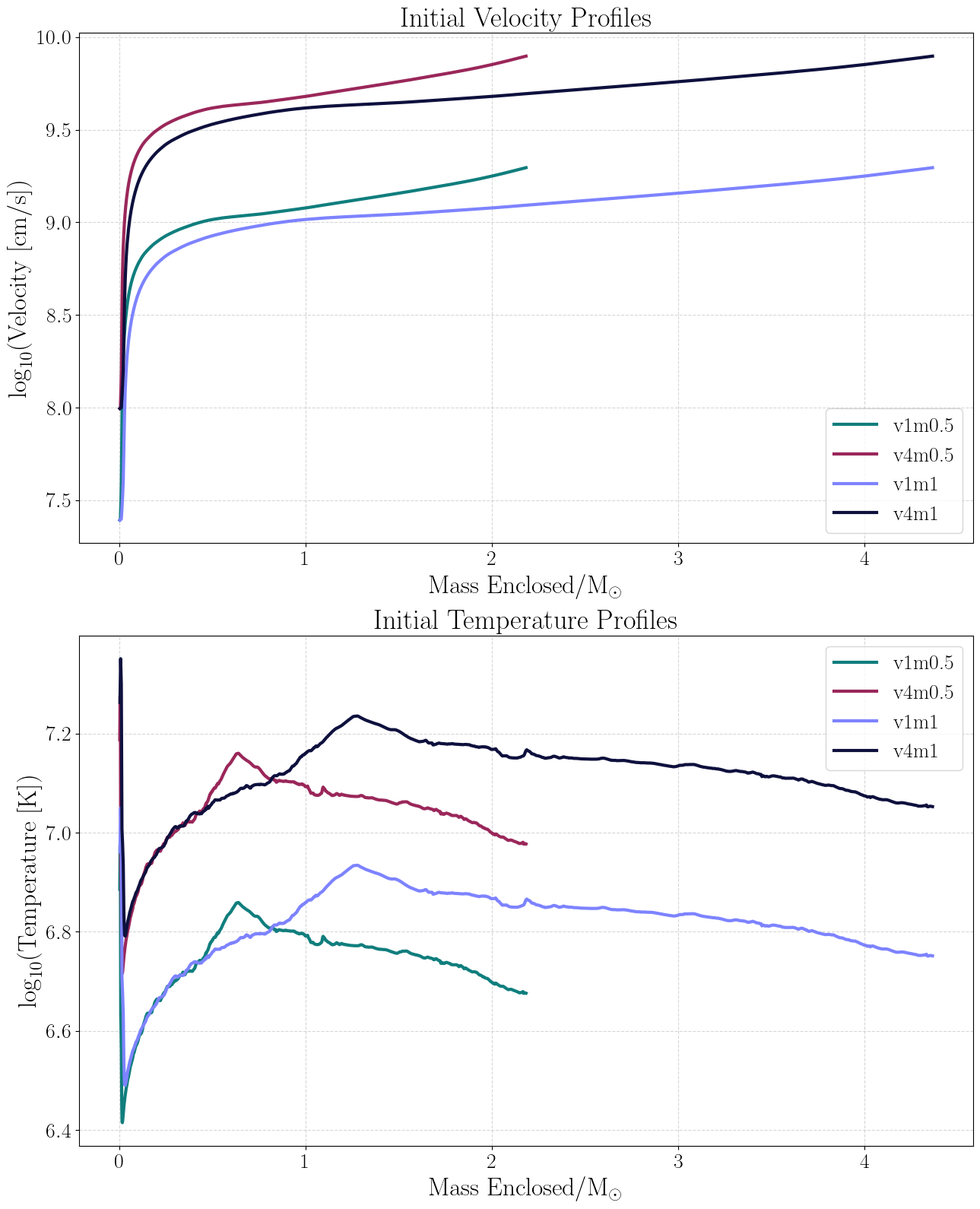}
 \caption{Initial velocity and temperature profiles for a sampling of our models. We maintain the same profile as the M25aE7.42 from \cite{2018ApJ...856...63F} and scale up the velocity. The temperature scales with the density (mass) and velocity to mimic internal shock-heating.}
 \label{fig:initial}
\end{figure}

The velocity profile (shown in Figure~\ref{fig:initial}) can dramatically alter the light-curve. For example, with interior energy sources (magnetar, fallback, unmixed $^{56}$Ni), the amount of slowly moving ejecta can drastically change how quickly this energy can diffuse out to the photosphere. In our model, the innermost $\sim 2.4\,M_\odot$ of ejecta is moving very slowly at this time and we expect it to fallback~\citep{2009ApJ...699..409F}. With this material accreting onto the neutron star, the total ejecta mass for this explosion is $4.39\,M_\odot$ For our standard model, we assume this material has fallen back onto the neutron star and is accreted before it can significantly alter the supernova emission. Were we to include it, the rise time from any of our central energy sources would be significantly delayed.

Like many hydrodynamics explosions of supernovae, \cite{2018ApJ...856...63F} did not sufficiently resolve the edge of the C/O star to capture the potential acceleration of a small amount of relativistic material expected as the shock breaks out of the star~\citep{2001ApJ...551..946T}. But we do include the internal shocks (these would be stronger in 3-dimensional explosions) that can power an initial shock emergence signal. The strength of these shocks produces the initial temperature profile (shown in Figure~\ref{fig:initial}). For compact stars, this temperature drops dramatically (for adiabatic expansion, $T \propto \rho^{1/3} \propto {\rm radius}^{-1} \propto {\rm temperature}^{-1}$). By the time the star expands to become large enough to drive a strong light-curve, the temperature from internal shocks is too low to drive emission.
 
Rather than use a broad set of simulated progenitors and energies, we scale this base model to produce a range of explosions, varying the velocities, initial masses of $^{56}$Ni, nickel mixing, and mass. Different parameter combinations are given in Table \ref{table-param}. The naming convention we use for the duration of this paper follows this format: v1ni0.2f1m1. `v' refers to velocity and is followed by the multiplier for the initial velocity of each zone. `ni' refers to Ni$^{56}$ and is followed by the initial nickel mass in M$_\odot$. `f' gives the percentage of each zone, from the center outwards, composed of nickel at the onset of the simulation. For example, f0.25 refers to the fact that a quarter of each zone mass starting from the inner-most zone was set to be Ni$^{56}$ until the target nickel content has been reached. Finally, `m' gives the pre-simulation mass multiplier. We consider our baseline model to be v1ni0.2f1m1; velocity multiplier one, $0.2M_\odot$ of Ni$^{56}$ all centrally located, and mass multiplier one. By using this scaling approach, we can better disentangle the effects each of these quantities on the early supernova light-curve. In this section, we describe our choice of parameter studies.

\begin{table}[!ht]
\begin{center}
\begin{tabular}{l|cccc}
\hline\hline 
& \textbf{ni0} & \textbf{f1} & \textbf{f0.25} & \textbf{f0.1} \\
\hline
\textbf{v1m0.5} & ni0 & ni0.1, 0.2, 0.4 & ni0.2 & ni0.1, 0.2, 0.4 \\
\textbf{v1m1} & ni0 & ni0.1, 0.2, 0.4 & ni0.2 & ni0.1, 0.2, 0.4 \\
\textbf{v2m0.5} & ni0 & ni0.2 & ni0.2 & ni0.2 \\
\textbf{v2m1} & ni0 & ni0.2 & ni0.2 & ni0.2 \\
\textbf{v4m0.5} & ni0 & ni0.2 & ni0.2 & ni0.2 \\
\textbf{v4m1} & ni0 & ni0.2 & ni0.2 & ni0.2 \\
\textbf{v8m0.5} & ni0 & ni0.2 & ni0.2 & ni0.2 \\
\textbf{v8m1} & ni0 & ni0.2 & ni0.2 & ni0.2 \\
\hline
\end{tabular}
\caption{Initial conditions for nickel powered SNe.}
\label{table-param}
\end{center}
\end{table}

\subsection{Mass}

The ejecta mass is determined by the mass of the C/O core and the amount of material that accretes onto the compact object prior to the explosion. The expected masses depend upon the engine driving Ic-BL supernovae. The two dominant engines ~\citep{menegazzi2024varietydiscwinddrivenexplosions} for Ic-BL are the the black hole accretion disk, e.g. collapsar~\citep{1993ApJ...405..273W}, and magnetar engines~\citep{2000ApJ...537..810W,2001ApJ...552L..35Z}. The ejecta masses expected from these engines are dictated by two competing effects: the required progenitor mass and the remnant mass. Black hole formation (needed for black hole accretion disk engines) typically occurs in more massive stars~\citep{1999ApJ...522..413F}. These stars produce more massive C/O cores, but more of the C/O core mass accretes onto the compact remnant to form a $\sim 3\,M_\odot$ black hole. In contrast, magnetar engines typically have low-mass C/O cores but the remnant mass is closer to $\sim 1.4\,M_\odot$. Our 25\,M$_\odot$ model has a roughly $3.5\,M_\odot$ compact remnant yielding $4.39\,M_\odot$ of ejecta. This progenitor mass is nearly at the edge of black hole versus neutron star forming stars~\citep{1999ApJ...522..413F}. By varying this mass by a scaling factor ($f_{\rm mass}$), we capture the rough features of the different engines: magnetar model ($0.5<f_{\rm mass}<1.5$) and black hole accretion disk model ($1.0<f_{\rm mass}<4$). 

We note that, for Ic-BL, the black hole accretion disk model recludes the formation of events at high-metallicity~\citep{fryer2024explainingnonmergergammaraybursts}, matching the current low-metallicity environments of observed Ic-BL supernovae~\citep{Modjaz_2008}.  At this time, this engine scenario is the leading candidate to explain Ic-BL SNe with a corresponding mass range of:  $1.0<f_{\rm mass}<4$. 

\subsection{Velocity and Energetics}

For the asymmetric explosions in our Ic-BL engines, we may be able to produce a broad range of explosion energies. To probe the effects of explosion energy, we scale the ejecta velocities. We preserve our velocity profile, scaling it through a constant factor ($f_{\rm vel}$) of 1, 2, 4, or 8. These velocity values correspond to spherical explosion energies of 7.45e51, 2.98e52, 1.19e53, and 4.77e53 ergs respectively. The higher 2 energies are almost certainly beyond the total energy of Ic-BL supernovae, but studying their evolution provides insight into the physics of the supernova explosion. The explosion velocity both alters the time-dependence of the energy diffusion and time evolution of the photosphere (which effects the photospheric radius). For shock-heating models, it changes the amount of kinetic energy available to power our supernova emission.


\subsection{Temperature}

For our models, we assume the initial temperature will also be altered by the variation in the mass and velocity of our explosions. To determine this temperature, we assume the initial temperature is set by the shocks produced as the blast wave propagates through the star. We assume strong shock jump conditions~\citep{1989ApJ...346..847C} and a radiation pressure dominated gas. For the energetic explosions of our Ic-BL models, both assumptions are valid. What our 1-dimensional progenitor model misses is the ancillary shocks that would further heat this ejecta. Our standard model increases the temperature by 30\% (this value was chosen in an adhoc matter loosely based on the results from multi-dimensional explosion models~\citep{2020ApJ...895...82V}) to include these ancillary shocks.

We assume that varying the mass and velocity also affects the nature of the heating in these internal shocks. Under our strong shock solution assuming radiation-pressure dominated gas, the temperature scaling is a function of our choice for velocity and mass~\citep{Fryer_2020}:
\begin{equation}
 T_{\rm shock} \propto \rho_{\rm star}^{1/4} v_{\rm shock}^{1/2}
\end{equation}
corresponding to a temperature scaling set by the mass and velocity scaling:
\begin{equation}
 f_{temp} \propto f_{\rm mass}^{1/4} f_{\rm vel}^{1/2}.
\end{equation}
If we increase the velocity or mass, the initial temperature will increase as well, causing a brighter initial shock breakout signal. But because our progenitor is so compact, this only affects the first few days of the supernova emission.  A key aspect of this derivation is not that the gas is radiation-pressure dominated, but that the ions, electrons, and photons achieve some rough equilibrium.  The shock energy is input into the ions and electron/ion coupling (reasonably accurate for the light-curve at the early-times studied here) sets the electrons to the same temperature as the ions.  However we also assume that the electron and radiation temperatures are in rough equilibrium.  In temperature equilibrium, the energy from the shock will predominantly lie in the radiation.  If temperature equilibrium is not achieved, the shock emission will have a higher-temperature spectrum than that assumed in these models.  The existence of high energy emission in the ultraviolet is a clear sign of shock interactions.  We will discuss this further in our results section.

\subsection{Ni source}

For Ic-BL explosions, $^{56}$Ni is the dominant radioactive decay power source. For rotating models with accretion disks, $^{56}$Ni can be produced two ways: explosive nucleosynthesis as the shock propagates through the star~\citep{2001ApJ...555..880N} and in the accretion disk itself~\citep{2006ApJ...643.1057S}. If the former is the dominant $^{56}$Ni production site, the amount produced depends upon the structure of the silicon layer as well as the shock energy. More nickel will be produced in stronger explosions with compact cores (more massive stars). In scenarios where the engine occurs after a normal or weak supernova explosion, the total $^{56}$Ni yield will typically be low (below $0.05$\,M$_\odot$) and may even be negligible~\citep{2006ApJ...650.1028F}. In the latter case where disk winds contribute to the $^{56}$Ni production, the amount of $^{56}$Ni produced in the disk depends both on the accretion rate and the amount of mass in the disk. For accretion rates lying between 0.1-1\,M$_\odot \, {\rm s^{-1}}$, there is very little deleptonization in the disk and most of the disk wind will be $^{56}$Ni~\citep{2006ApJ...643.1057S}. These accretion rates are expected for most stellar progenitors~\citep{fryer2024explainingnonmergergammaraybursts}. In this case, the amount of $^{56}$Ni yield depends primarily on the amount of material that accretes through the disk and the amount of material ejected in the wind. For instance, if 20\% of the disk is ejected in the wind and $2\,M_\odot$ passes through the disk, the $^{56}$Ni production will be $0.4\,M_\odot$. For our particular model, this is roughly an upper limit on $^{56}$Ni production. However, more massive stars can produce much more and lower mass stars will produce much less.

Although observations have been used to estimate the $^{56}$Ni yield, most of those studies assume the peak luminosity is powered by $^{56}$Ni decay~\citep{Anand_2024, refId0}. As we shall see in this paper, this assumption is unlikely to be true and over-estimates the $^{56}$Ni yield from these explosions. Indeed, there are observations showing that the inferred $^{56}$Ni is higher than the more reliable late time observations~\citep{Cano2017}. 

Another important factor in the radioactive power source is the distribution of the $^{56}$Ni. $^{56}$Ni is produced in the innermost ejecta and the energy deposited in its decay must diffuse out from its position to affect the observed emission. Where the $^{56}$Ni is located dictates when its decay energy can reach the photosphere. Asymmetries in the explosion can mix the $^{56}$Ni out in the star and many observations require extensive mixing to match observations. 

To investigate the effect of mixing and nickel mass on the light-curves of Ib/c supernovae, we perform a variety of simulations powered exclusively by nickel decay. Initial nickel masses included 0, 0.1, 0.2, and 0.4 M$_\odot$. These values encompass the range derived in \citet{refId0}. $^{56}$Ni was either placed at the center of the star or mixed outwards to allow it to reach the photosphere more quickly. To ``mix out" the nickel, a percentage of the mass for each zone (10\%, 25\%, or 100\%) was initialized as $^{56}$Ni until the total nickel mass reached our desired value. These parameters are adjusted in this paper in an effort to create a viable rise time, fall time, shape, and peak luminosity for the resulting light-curve.

\subsection{shock-heating}

Given both the complex binary progenitors required for Ic-BL supernovae (multiple binary shell ejecta, strong Wolf-Rayet winds), shock interactions at some level must occur. For binary mass ejection, it depends on when they occur, anywhere between 7-700\,d depending on the engine and progenitor~\citep{fryer2024explainingnonmergergammaraybursts}. For clumpy winds, the strongest interactions are likely to occur at early times (e.g. immediately after shock breakout). 

Evidence of shock interactions occurs in a wide range of supernovae~\citep{refId0,2017hsn..book..403S,2019ApJ...874...68C}, and the most obvious examples show strong shell interactions, e.g. the late-time peak of SN 2022xxf~\citep{refId0}. The nature of the circumstellar interactions is not well known. Only preliminary radiation-hydrodynamics calculations have been done modeling the nature of clumpy winds~\citep{2021ApJ...910...68J} and much more work is necessary to determine how their interactions affect the supernova light-curves~\citep{Fryer_2020}. Most shell interaction studies have assumed spherically-symmetric shells, but it is likely that these shells also have complex structures (e.g. porosity) that complicate the shock interactions.

Our two shock-interaction prescriptions (Equations~\ref{eq:vshock},~\ref{eq-sig}) allow us to better capture the wide range of conditions from these interactions. We have a number of free parameters: the beginning and end of the shock interactions, the amount of velocity deceleration in the shock interactions, and, for the sigmoid function, the evolution of the shock interactions (steepness). We perform two sets of shock-heating models. For our sigmoid models, we adjust the shock-heating duration, onset time, and steepness to develop models that fit the single peak Ic-BL supernovae. We also include a suite of strong shock models that drive shocks at different times to produce a broader range of potential shock-heating solutions. We would like to stress that, at this time, the uncertainties in shock interactions is sufficient that we can fit a broad range of data. Initial conditions for shock-powered runs are given in Tables \ref{table-shock},\ref{table-strongshock}.

\begin{table}[!ht]
\begin{center}
\begin{tabular}{l|cccc}
\toprule
\toprule
\textbf{Shock Start} & \textbf{4h} & \textbf{4h} & \textbf{0} \\
\textbf{Shock End} & \textbf{25d} & \textbf{25d} & \textbf{35d} \\
\textbf{Final Velocity} & \textbf{98\%} & \textbf{96\%} & \textbf{95\%} \\
\hline
\textbf{v1m0.5} & ni0 & ni0 & ni0\\
\textbf{v1f0.25m0.5} & ni0.2 & ni0.2 & ni0.2 \\
\textbf{v1f0.25m1} & ni0.05, 0.2 & ni0.05, 0.2 & ni0.05, 0.2 \\
\textbf{v1f1m0.5} & ni0.2 & ni0.2 & ni0.2 \\
\textbf{v1f1m1} & ni0.05, 0.2 & ni0.05, 0.2 & ni0.05, 0.2 \\
\bottomrule
\end{tabular}
\caption{Initial conditions for shock powered SNe where the shock is assumed to be gradual based on clumpy winds and the shock start time is within 4 hours of breakout and the end times are varied from 25-35d.}
\label{table-shock}
\end{center}
\end{table}

\begin{table}[!ht]
\begin{center}
\begin{tabular}{l|cccc}
\toprule
\toprule
\textbf{$f_{\rm mass}=1$}, \textbf{$M_{^{56}Ni}=0$} &  & & \\
\hline
Model (sh) & $f_{\rm vel}$ & $t^0_{\rm shock}$ & $t^{\rm end}_{\rm shock}$ & $(v-v_{\rm fin})/v$ \\
\hline
\textbf{v1m1t5t10v0.04} & 1 & 5d & 10d & 0.04 \\
\textbf{v2m1t5t10v0.04} & 2 & 5d & 10d & 0.04 \\
\textbf{v2m1t5t10v0.08} & 2 & 5d & 10d & 0.08 \\
\textbf{v2m1t20t25v0.04} & 2 & 20d & 25d & 0.04 \\
\textbf{v4m1t15t15.25v0.04} & 4 & 15d & 15.25d & 0.04 \\
\textbf{v4m1t15t16v0.04} & 4 & 15d & 16d & 0.04 \\
\textbf{v4m1t20t25v0.04} & 4 & 20d & 25d & 0.04 \\
\bottomrule
\end{tabular}
\caption{Initial conditions for shock powered SNe where shock is assumed to be powered by a porous shell interaction (we focus on the earliest times expected from~\cite{fryer2024explainingnonmergergammaraybursts}.}
\label{table-strongshock}
\end{center}
\end{table}

\section{Results}
\label{sec:results}

Our grid of models allows us to better understand how the ejecta properties and the energy sources drive a supernova light-curve. In this section, we present the results of this grid, focusing on the effects of the mass, velocity, and nickel distribution on the bolometric light-curves. We compare these to models powered by shock interactions where we focus our smoothed (sigmoid) interaction models to mimic the peak light-curve of Ic-BL supernovae and use our strong shocks to study the broader potential of shock-heating. 

Before we study these trends (both from radioactive decay and shock-heating), we first use our simulations to provide some insight into the nature of an explosive light-curve. We will use this insight to understand the trends in our explosion models.

\subsection{Anatomy of a Light-Curve}

The luminosity of a thermal Bremsstrahlung-emitting object is roughly fit by a blackbody and is proportional to both its temperature and emitting area. For spherically expanding ejecta, this luminosity is:
\begin{equation}
 L_{\rm therm} = 4 \pi r^2_{\rm phot} \sigma T^4_{\rm phot}
 \label{eq:bb}
\end{equation}
where $r_{\rm phot}$, $T_{\rm phot}$ are the photospheric radius and temperature respectively and $\sigma = 5.67 \times 10^{-5} {\rm erg \, s^{-1} \, cm^{-2} \, K^{-4}}$ is the Stefan-Boltzmann constant. We will use this simple emission picture to build our intuition of the supernova light-curve~\citep[also see][]{1982ApJ...253..785A,fryer224}.  Using the the mass, velocity and opacity of the exploding material, one can derive a simple parameter ($y$) dictating the shape of the light-curve as a function of the ejecta mass ($M_{\rm exp}$) and explosion energy or velocity ($v_{\rm exp}$) ~\citep{1982ApJ...253..785A,2000ApJ...530..744P,2017ApJ...846...33A}:
\begin{equation}
    y = (\kappa_t M_{\rm exp}/v_{\rm exp})^{1/2}
\end{equation}
where $\kappa_t$ is the opacity.  However, we caution against over-interpreting these simplified models which make a number of approximations that are not strictly valid. For example, the photosphere is not a single radius. For asymmetric explosions, the photosphere will vary as a function of angle and this can cause considerable variation in the light-curve. In addition, even with a spherical explosion, the photosphere is wavelength dependent. Both of these effects have been shown to complicate the interpretation of kilonovae observations~\citep{2021ApJ...910..116K,2024ApJ...961....9F}. Nevertheless, the toy picture of a supernova where the ejecta expands (evolving the photospheric radius) and cools (both through emission and adiabatic expansion) provides considerable insight into the trends with explosion properties and power sources in the light-curve.

\begin{figure}
 \centering
 \includegraphics[width=\linewidth]{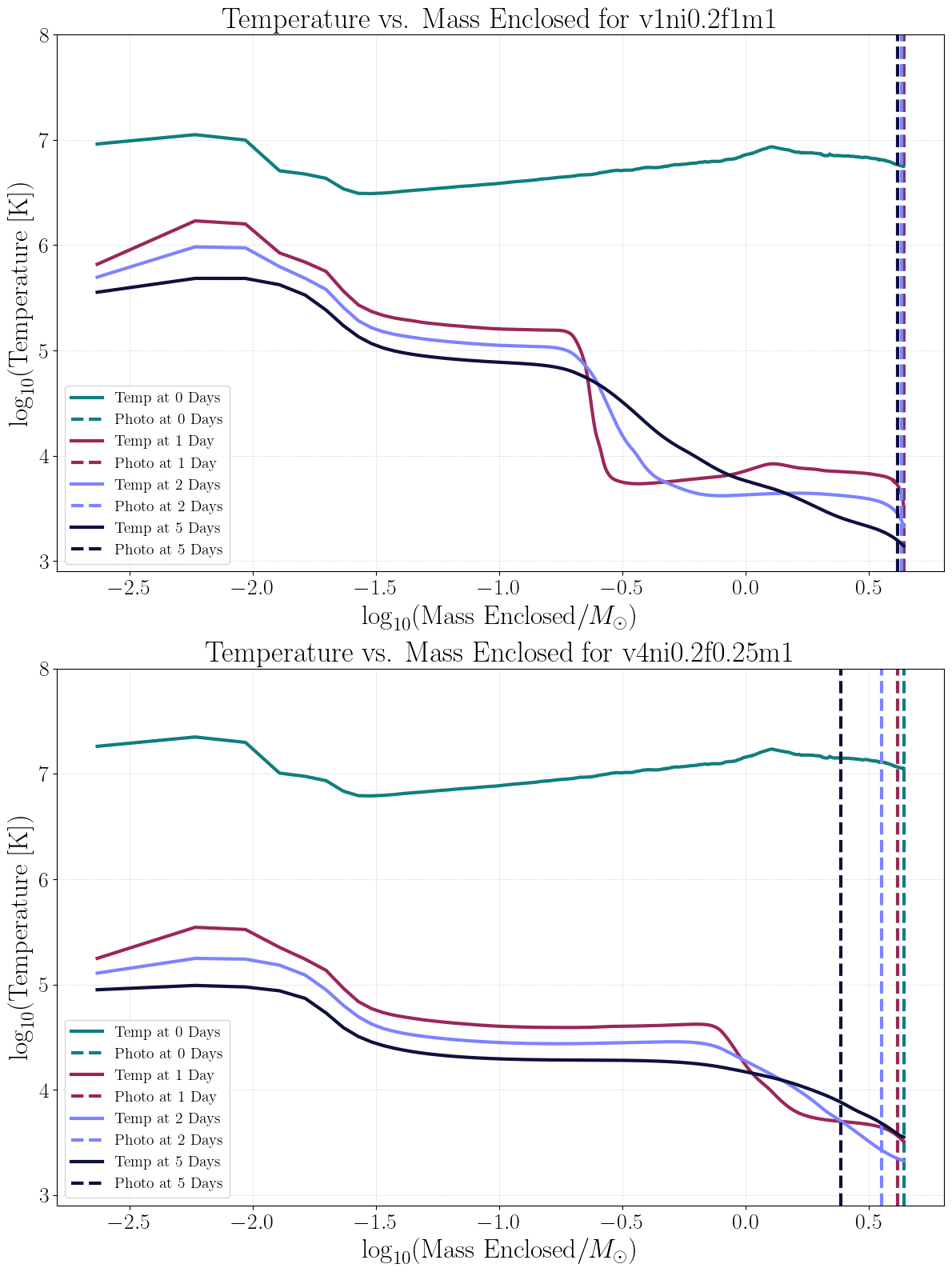}
 \caption{Temperature profiles at 0, 1, 2, and 5 days for two nickel powered simulations. The rapid cooling from expansion occurs in the first day. The position of the strong temperature gradient (enclosed mass of 0.3\,M$_\odot$ for the $f_{\rm mix}=1$ case and $\sim 1$,M$_\odot$ for the $f_{\rm mix}=0.25$ case) shows the rough position of the $^{56}$Ni decay. With time, this energy begins to diffuse to the photosphere.}
 \label{fig:temp}
\end{figure}

Figure~\ref{fig:temp} shows the temperature profile of two of our models at four different times in the explosion with vertical lines denoting the gray opacity photosphere (determined by the $\tau = 1$ surface). The initial temperature at breakout can exceed $10^6$\,K, emitting in the ultraviolet and X-ray, but the radius of these C/O cores is very compact and this burst of X-rays will be fleeting. As the ejecta expands, it cools adiabatically ($T_{\rm photo} \propto r_{\rm photo}^{-1}$). In homologous expansion at early times, the photosphere follows the ejecta:
\begin{equation}
 r_{\rm photo}^{-1} = v_{\rm shock} t_{\rm shock}
\end{equation}
where $v_{\rm shock},t_{\rm shock}$ are the velocity and time evolution of the shock respectively. By 1 day, the ejecta has expanded by over a factor of 1000 and the temperature has dropped a corresponding amount. The early-time emission from internal shock will drop dramatically: 
\begin{equation}
 L_{\rm therm} r^2_{\rm phot} T^4_{\rm phot} \propto r^{-2} \propto t_{\rm shock}^{-2}
\end{equation}
where we have assumed adiabatic and homologous expansion with $t_{\rm shock}$ as the shock evolution timescale. In addition, the outer material cools through emission. After this temperature drop, some power source is required to drive a luminous supernovae.

The energy from the inner part of the star diffuses outward, reheating the material at the photosphere. The top panel of Figure~\ref{fig:temp} shows the evolution of a model where the $^{56}$Ni is concentrated in the innermost ejecta (where it is initially produced). The temperature profile shows the energy deposited in $^{56}$Ni decay diffusing out of the star. When this energy reaches the photosphere, it starts to power the light-curve. If the $^{56}$Ni is mixed out (bottom panel), its decay energy has less material to diffuse through and it will power the observed emission more quickly.

\begin{figure}
 \centering
 \includegraphics[width=\linewidth]{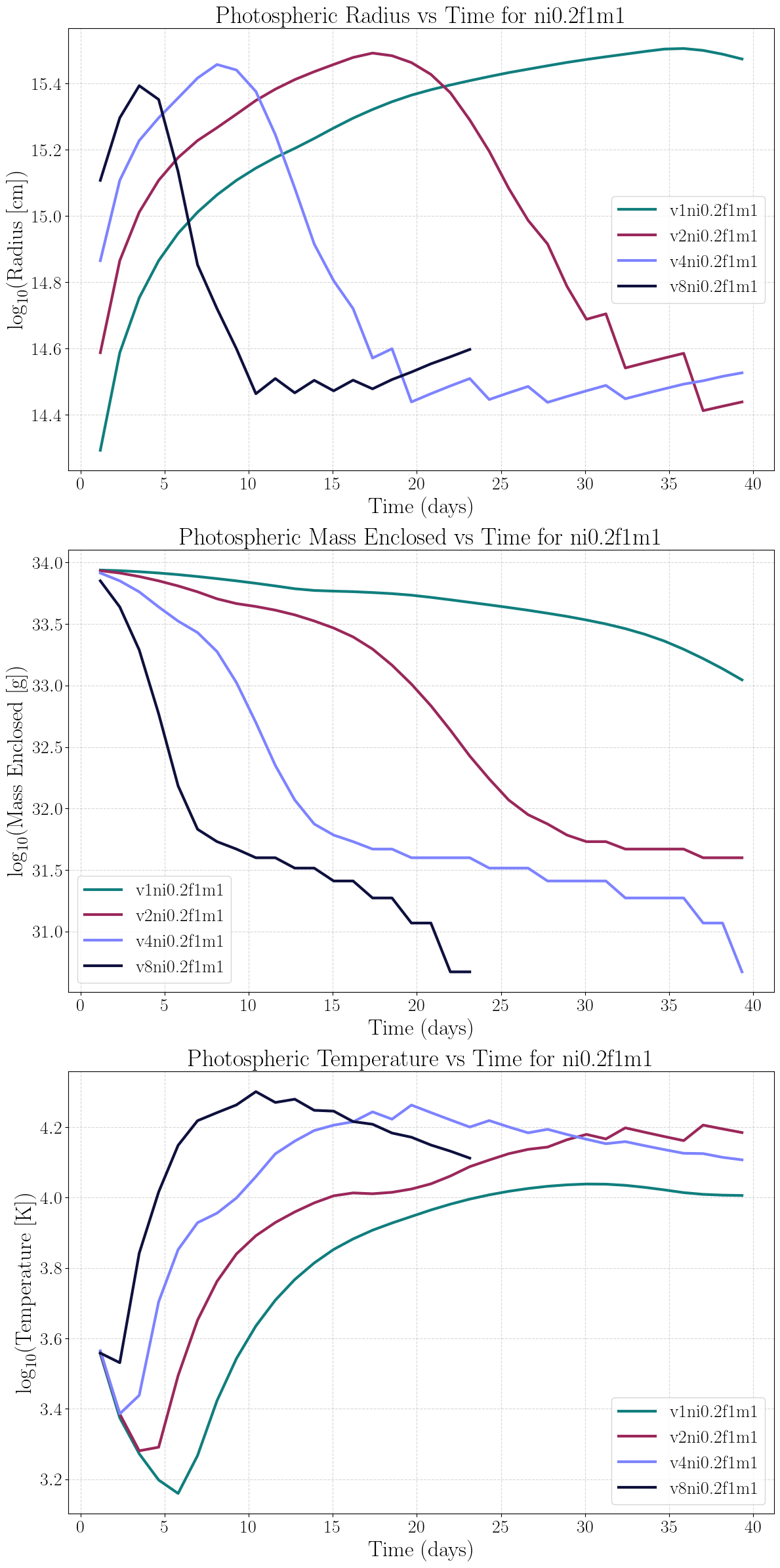}
 \caption{Photospheric properties for the ni0.2f1m1 models varying $f_{\rm vel}$ from 1 to 8. The black line stops when the photosphere has reached the center of the star.}
 \label{fig:photo}
\end{figure}

At the same time that the decay energy is diffusing outward, the photosphere moves inward in mass as the outer layers of the star become optically thin. As the photosphere meets the diffusion energy from nickel decay, the luminosity will increase greatly. The photosphere always moves inward in mass space.  However, since the ejecta is moving outward, the radius of the photosphere can increase. The motion of the photosphere in both mass and radius for a sampling of our models is shown in Figure~\ref{fig:photo}. For our standard `v1' models, the photosphere only begins to move inward in radius after 35\,d. But the time of the maximum radial extent of the photosphere is inversely proportional to the velocity (this is because the radial expansion of the star is velocity times time, so the extent of the `v8' model at a given time is the same as the `v1' model at 8 times that time). For the `v8' models, this peak occurs within 5 days of the explosion.

For our nominal model, the photosphere is still within the star at 40\,d, but for our other models, the photosphere ultimately reaches the center of the ejecta. This is where our simple luminosity formula (Equation~\ref{eq:bb}) breaks down. Just because the optical depth is less than one doesn't mean that there is no interaction between the radiation and the matter, and capturing this interaction is still important in light-curve calculations (most transport calculations capture this but analytic approaches will not). In addition, when the photosphere reaches the center of the ejecta depends sensitively on our choice of the lowest ejecta velocity. It may be that a compact remnant is obscured for years after the explosion~\citep{1999ApJ...511..885F}. We assumed most of our low velocity ejecta would fall back into our supernovae. If we had included the low velocity ejecta, most of our $^{56}$Ni energy would be trapped well beyond the peak of emission.

With this basic supernova understanding, let us discuss the trends in our light-curves for the different power sources.

\subsection{Powered by Ni decay}

Although many studies assume that the light-curve of type Ib/c supernovae are powered by $^{56}$Ni decay, such models require that the energy released in radioactive decay can reach the photosphere to power the observed emission. In this section, we study the effects produced by varying the ejecta properties (velocity, mass, $^{56}$Ni mass and its distribution). Table~\ref{table_nickel} shows the peak luminosities plus half-peak rise and decay times for our grid of models. In this section, we discuss the trends in this grid of models.

\begin{table}[ht]
\centering
\begin{tabular}{lcccc} 
\toprule
\toprule
\textbf{Parameters} & \textbf{log$_{10}$(L$_{\odot, p})$} & \textbf{t$_{p}$} & \textbf{t$_{-1/2}$} & \textbf{t$_{+1/2}$} \\
\midrule
v1ni0.2f0.1m0.5 & 42.8 & 15.6 & 8.9 & 22.0 \\
v2ni0.2f0.1m0.5 & 42.7 & 9.8 & 5.8 & 17.6 \\
v4ni0.2f0.1m0.5 & 42.5 & 5.9 & 3.7 & 14.8 \\
v8ni0.2f0.1m0.5 & 42.2 & 3.4 & 2.3 & 11.8 \\
v1ni0.2f0.1m1 & 42.7 & 29.6 & 14.2 & 31.5 \\
v2ni0.2f0.1m1 & 42.5 & 18.8 & 9.1 & 29.6 \\
v4ni0.2f0.1m1 & 42.3 & 12.7 & 6.7 & 31.0 \\
v8ni0.2f0.1m1 & 41.9 & 9.7 & 5.9 & 32.2 \\
v1ni0.2f0.25m0.5 & 42.7 & 25.2 & 11.8 & 25.2 \\
v2ni0.2f0.25m0.5 & 42.6 & 16.3 & 7.9 & 26.0 \\
v4ni0.2f0.25m0.5 & 42.3 & 11.3 & 6.1 & 27.9 \\
v8ni0.2f0.25m0.5 & 41.9 & 8.8 & 5.4 & 29.5 \\
v1ni0.2f0.25m1 & 42.7 & 35.2 & 14.2 & 32.5 \\
v2ni0.2f0.25m1 & 42.5 & 22.9 & 9.7 & 35.4 \\
v4ni0.2f0.25m1 & 42.1 & 17.0 & 8.6 & 45.1 \\
v8ni0.2f0.25m1 & 41.7 & 14.9 & 9.3 & 55.1 \\
v1ni0.2f1m0.5 & 42.7 & 29.1 & 10.5 & 26.6 \\
v2ni0.2f1m0.5 & 42.5 & 19.7 & 8.0 & 34.5 \\
v4ni0.2f1m0.5 & 42.1 & 16.0 & 8.6 & 48.0 \\
v8ni0.2f1m0.5 & 41.5 & 15.5 & 10.3 & $>$64.5 \\
v1ni0.2f1m1 & 42.6 & 37.8 & 12.4 & 34.6 \\
v2ni0.2f1m1 & 42.4 & 26.4 & 10.3 & 45.5 \\
v4ni0.2f1m1 & 41.9 & 23.6 & 13.2 & $>$56.4 \\
v8ni0.2f1m1 & 41.3 & 26.6 & 18.8 & $>$53.4 \\
v1ni0.1f0.1m0.5 & 42.4 & 21.9 & 10.5 & 22.5 \\
v1ni0.4f0.1m0.5 & 42.8 & 15.2 & 9.7 & 21.8 \\
v1ni0.1f0.1m1 & 42.4 & 31.5 & 13.1 & 24.3 \\
v1ni0.4f0.1m1 & 43.0 & 20.9 & 12.0 & 26.6 \\
v1ni0.1f1m0.5 & 42.4 & 27.5 & 9.0 & 24.7\\
v1ni0.4f1m0.5 & 43.0 & 29.9 & 12.2 & 24.3 \\
v1ni0.1f1m1 & 42.3 & 35.8 & 10.8 & 29.2 \\
v1ni0.4f1m1 & 42.9 & 39.8 & 14.4 & 27.2 \\
v1ni0m0.5 & 38.2 & 0.5 & 0.1 & 2.5 \\
v2ni0m0.5 & 38.8 & 0.5 & 0.2 & 2.1 \\
v4ni0m0.5 & 39.4 & 0.3 & 0.2 & 1.7 \\
v8ni0m0.5 & 40.1 & 0.1 & - & 0.8 \\
v1ni0m1 & 38.4 & 0.6 & 0.1 & 2.7 \\
v2ni0m1 & 39.1 & 0.3 & 0.1 & 1.6 \\
v4ni0m1 & 39.6 & 0.5 & 0.3 & 2.1 \\
v8ni0m1 & 40.2 & 0.2 & - & 1.4 \\

\bottomrule
\end{tabular}
\label{table_nickel}
\caption{Peak luminosity, time of peak, time from L$_p$/2 to L$_p$, and time from L$_p$ back to L$_p$/2. A dash in the t$_{-1/2}$ column indicates that the brightness never increased during the simulation. This is only true for runs with no nickel.}
\end{table}

\begin{figure}
 \centering
 \includegraphics[width=\linewidth]{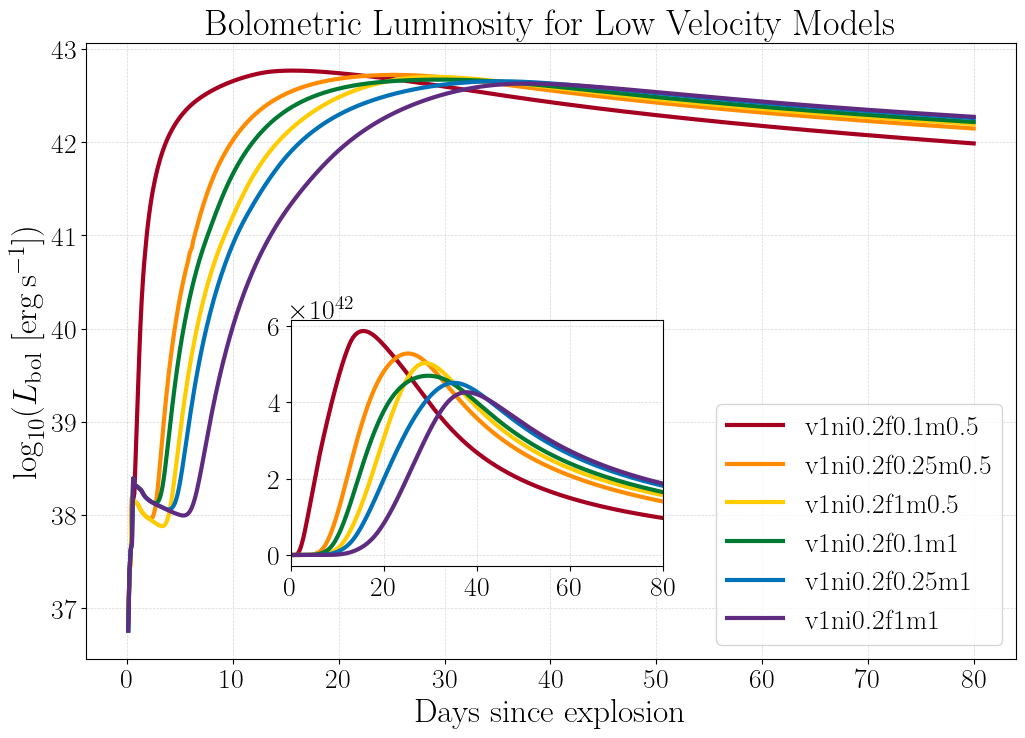}
 \caption{Bolometric light-curves for a subset of $f_{\rm vel} = 1.0$ models using two different values for $f_{\rm mass} (0.5, 1.0)$ and 3 different mixing parameters: $f_{\rm mix} = 1, 0.25, 0.1$. For $f_{\rm mix} = 1$, the $^{56}$Ni is located in the innermost ejecta. For the $f_{\rm mass} = 0.5, f_{\rm mix}=0.1$ model, it is nearly mixed out to the outer edge. The inset plot shows the same data without the logarithmic y-axis.}
 \label{fig:ni_lc}
\end{figure}

Figure~\ref{fig:ni_lc} shows the bolometric light-curves for a subset of our $f_{\rm vel} = 1.0$ models with $f_{\rm mass} = 0.5, 1.0$, total $^{56}$Ni mass of $0.2\,M_\odot$ and 3 different values for the mixing of this nickel: $f_{\rm mix} = 1, 0.25, 0.1$ corresponding to no mixing and partial mixing out to 1 or 2\,$M_\odot$ in the ejecta. For the $f_{\rm mass} = 0.5, f_{\rm mix}=0.1$ models, $^{56}$Ni is mixed out almost to the edge of the ejecta. The purple curve demonstrates the delay in the role of radioactive decay if the radioactive isotope is limited to the interior where it is produced. In this light-curve, we see the initial emission powered by internal shock-heating. At roughly 7 days, the energy from $^{56}$Ni begins to heat the ejecta near the photosphere, but this light-curve doesn't peak until nearly 38\,d. If we mix the nickel outward or reduce the ejecta mass, the energy from radioactive decay can affect the light-curve more quickly. In the case of the $f_{\rm mass} = 0.5, f_{\rm mix}=0.1$, the contribution from the $^{56}$Ni is nearly immediate. In all of these models, the fall time of the light-curve is much longer than the rise time.

\begin{figure}
 \centering
 \includegraphics[width=\linewidth]{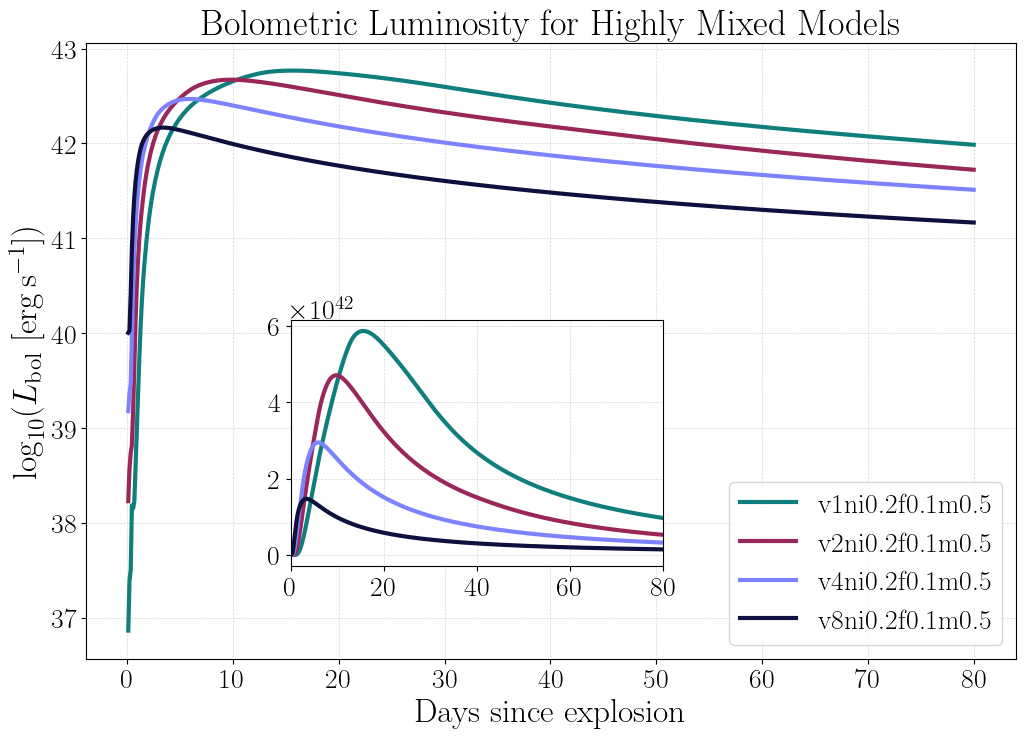}
 \caption{Bolometric light-curves for heavily mixed nickel powered models with four different velocity models: $f_{\rm vel} = 1,2,4 {\rm \, and }\, 8$. As the velocity increases, the light-curve peaks earlier, but is dimmer. The inset plot shows the same data without the logarithmic y-axis.}
 \label{fig:ext_lc}
\end{figure}

All else being equal, the peak emission in the light-curve for our $^{56}$Ni powered models depends linearly on the amount of energy sourced into the ejecta, i.e. the $^{56}$Ni mass (see Table~\ref{table_nickel}). But the velocity and mass variation can alter the peak luminosity as well and disentangling these properties may be difficult. The peak luminosity depends most sensitively on the explosion velocity (roughly linearly). Figure~\ref{fig:ext_lc} shows the behavior of the light-curve based on this velocity. Higher velocity models peak sooner, but at lower peak luminosities. The rapid rise of the light-curve for our fastest models limits the amount of $^{56}$Ni heating and, because the photosphere begins to move in rapidly, these models decay more quickly than slower models.

From Table~\ref{table_nickel}, we can infer other trends. The evolution of the emission around peak (particularly the rise and fall timescales) can be compared directly to observations to probe the explosion properties. As expected, the timescales are shorter for lower ejecta masses. Mixing can shorten the rise time, but it is less effective in decreasing the decay of the peak luminosity. As we shall see when we compare our simulations to the rise and fall times of observed Ic-BL supernovae, these $^{56}$Ni-dominated light-curves do not fit the existing data. It is possible to design models that better fit the observations (e.g. raising the lowest velocity in the ejecta, modifying the mixed distribution of the $^{56}$Ni such that it is primarily on the outside of the ejecta). But given these difficulties, it is worth considering alternative power sources for the peak emission.

\subsection{Strong Shocks from Shell Interactions}

Binary mass ejection (e.g. common envelope scenarios) will eject a shell of material. The ejection process will produce a shell with strong density variabilities (a.k.a. a porous shell). We mimic the supernova blast wave deceleration in this porous shell using Equation~\ref{eq:vshock}, which allows us to control the amount of deceleration as the shock propagates through the shell (typically we reduce the velocity between 4-8\%). In addition to our standard mass and velocity parameters (in these simulations we assume no $^{56}$Ni decay contributions), we can vary both the total deceleration and the interaction timescales (when the blast wave enters and exits the shell).

\begin{figure}
 \centering
 \includegraphics[width=\linewidth]{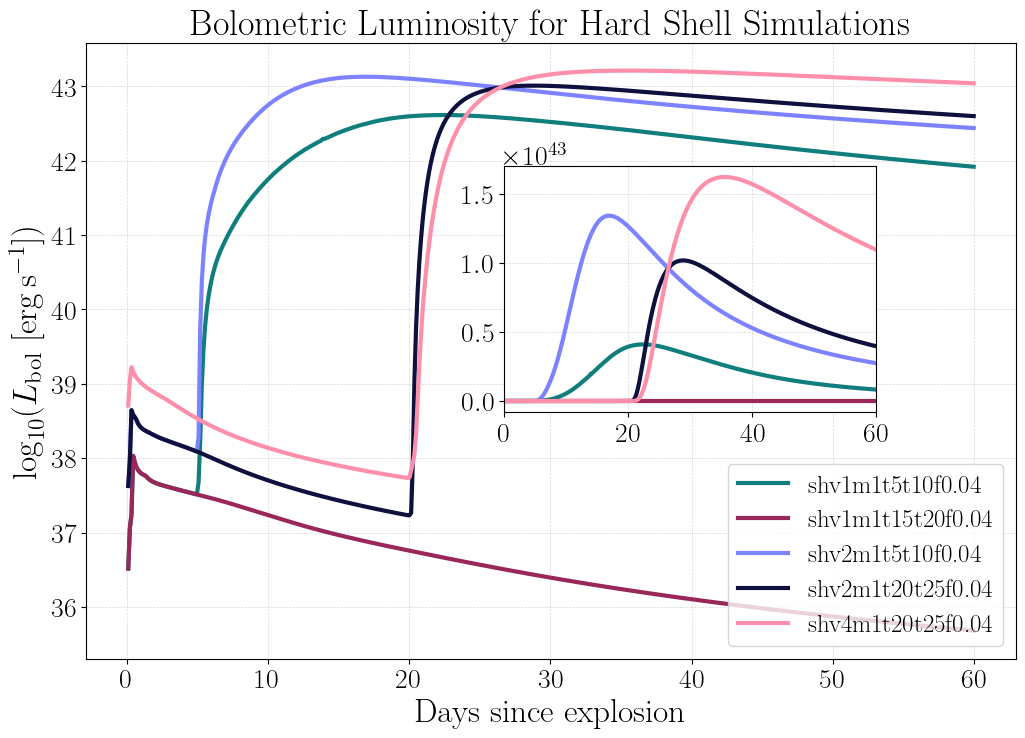}
 \caption{Bolometric light-curves of a sample of our shell interactions. These shell interactions are characterized by a sharp rise and a slower decay. The time of the rise is dictated by the position of the shell. The peak brightness depends upon the velocity of the shock and the amount of deceleration.}
 \label{fig:shellshock}
\end{figure}

Figure~\ref{fig:shellshock} shows a sampling of these calculations. Here we have focused on extremely nearby shells (e.g. shells within 0.01\,pc). This corresponds to shells ejected within a year of collapse: explosive mass loss events or helium common envelope ejections. The helium common envelope ejection process occurs in one proposed progenitor behind Ic-BL supernovae~\citep{fryer2024explainingnonmergergammaraybursts}. A feature of the strong shock implementation is that the rise time of the emission is sudden, occurring when the shock starts to decelerate. 

The duration of the emission is, in part, dictated by the duration of the interaction and we vary the end time of the interaction to get a better handle on this evolution. However, because the deceleration also sends a reverse shock through the ejecta, often well within the photosphere, the effect of this shock interaction on the light-curve lasts much longer than the shell interaction itself. This extension of the shock-driven luminosity will be much lower at late-times when the ejecta effected by the reverse shock is optically thin. Table~\ref{table-shock} shows the peak, rise, and fall timescales from our shock-driven models. Given the broad parameter space for these models, we can produce a wide range of results, meaning that such shell interactions are unlikely solutions to standard Ic-BL supernovae (unless there is a simple explanation for the creation of relatively uniform nearby shells). We argue that these shocks are more likely to explain peculiar events. Shocks with clumpy winds are much more likely to explain peak luminosities in Ic-BL supernovae.

\begin{table}[ht]
\centering
\begin{tabular}{lcccc} 
\toprule
\toprule
\textbf{Parameters} & \textbf{log$_{10}$(L$_{\odot, p})$} & \textbf{t$_{p}$} & \textbf{t$_{-1/2}$} & \textbf{t$_{+1/2}$} \\
\midrule
98\% v1ni0m0.5 & 42.3 & 15.6 & 4.4 & 7.2 \\
96\% v1ni0m0.5 & 42.8 & 17.7 & 6.4 & 6.5 \\
95\% v1ni0m0.5 & 42.9 & 23.1 & 7.4 & 8.4 \\
98\% v1ni0.2f0.25m0.5 & 42.9 & 23.3 & 9.8 & 9.0 \\ 
96\% v1ni0.2f0.25m0.5 & 43.0 & 19.7 & 8.2 & 15.4 \\
95\% v1ni0.2f0.25m0.5 & 43.3 & 23.1 & 5.6 & 6.5 \\
98\% v1ni0.2f1m0.5 & 42.8 & 25.9 & 10.9 & 21.8 \\
96\% v1ni0.2f1m0.5 & 42.9 & 24.8 & 12.7 & $>$15.2\\
95\% v1ni0.2f1m0.5 & 43.2 & 22.5 & 5.1 & 7.9 \\ 
98\% v1ni0.2f0.25m1 & 42.8 & 23.5 & 10.1 & 31.6\\ 
96\% v1ni0.2f0.25m1 & 43.1 & 20.3 & 9.8 & 12.6\\ 
95\% v1ni0.2f0.25m1 & 43.4 & 23.8 & 5.4 & 10.0\\ 
98\% v1ni0.2f1m1 & 42.8 & 36.3 & 12.3 & 19.4\\ 
96\% v1ni0.2f1m1 & 43.1 & 16.4 & 4.9 & 10.0\\ 
95\% v1ni0.2f1m1 & 43.3 & 25.2 & 9.5 & 11.2\\ 
98\% v1ni0.05f0.25m1 & 42.6 & 15.6 & 4.6 & 10.3\\
96\% v1ni0.05f0.25m1 & 43.1 & 14.7 & 4.3 & 10.5\\
95\% v1ni0.05f0.25m1 & 43.2 & 22.0 & 6.5 & 5.1\\
98\% v1ni0.05f1m1 & 42.6 & 15.6 & 4.7 & 8.9\\
96\% v1ni0.05f1m1 & 43.1 & 14.4 & 4.3 & 11.0\\
95\% v1ni0.05f1m1 & 43.2 & 21.2 & 5.7 & 11.5\\
Shell v1m1t5t10f0.04 & 42.6 & 22.5 & 8.2 & 17.6\\
Shell v1m1t15t20f0.04 & 38.0 & 0.5 & 0.1 & 1.2\\
Shell v2m1t5t10f0.04 & 43.1 & 16.9 & 6.4 & 17.4\\
Shell v2m1t20t25f0.04 & 43.0 & 28.8 & 5.7 & 22.7\\
Shell v2m1t5t10f0.08 & 43.1 & 17.0 & 6.4 & 17.1\\
Shell v4m1t15t15.25f0.04 & 43.3 & 26.6 & 7.3 & 30.1 \\
Shell v4m1t15t16f0.04 & 43.3 & 26.6 & 7.3 & 30.2 \\
Shell v4m1t20t25f0.04 & 43.2 & 35.5 & 9.8 & $>$24.5\\
\bottomrule
\end{tabular}
\label{table_shock}
\caption{Rise and fall times for shock-heated runs.}
\end{table}
 
\subsection{Smooth Shocks from a Clumpy Medium}

Our strong shock models mimic the luminosity we might expect from shock interactions with ejecta shells from binary mass ejecta. These shock interactions do not produce bolometric light-curve shapes that match typical peak emission in Ic-BL. But shock interactions with clumpy wind media are likely to have a different form. For these models, we use our sigmoid deceleration equation (Equation.~\ref{eq-sig}). Although this formalism also allows for a range of beginning and end times for the shock-heating, we expect these clumpy-wind shocks to occur immediately upon shock breakout. All of our models begin shock-heating within a few hours of explosion and this shock-heating dominates the emission from the onset of our models. 

Adjusting the shape parameter $k$ from Equation \ref{eq-sig} changes the time at which the velocity has decreased by half of the total desired value. By shifting this away from the mid point of the shock-heating period, we modify the shock-heating, altering the light-curves. Just as with the our strong shock models, we can control the magnitude of the shock-heating luminosity by varying the total deceleration of the velocity. Our shock-heating models that lower the velocity to 95\% of the original value over the course of $3 \times 10^6$\,s peak the highest at $>10^{43}$ ergs/s while the ones limited to 98\% of the total peak much lower at $10^{42}-10^{43}$ ergs/s. Control over this parameter makes it possible to simulate any peak luminosity. 

\begin{figure}
 \centering
 \includegraphics[width=\linewidth]{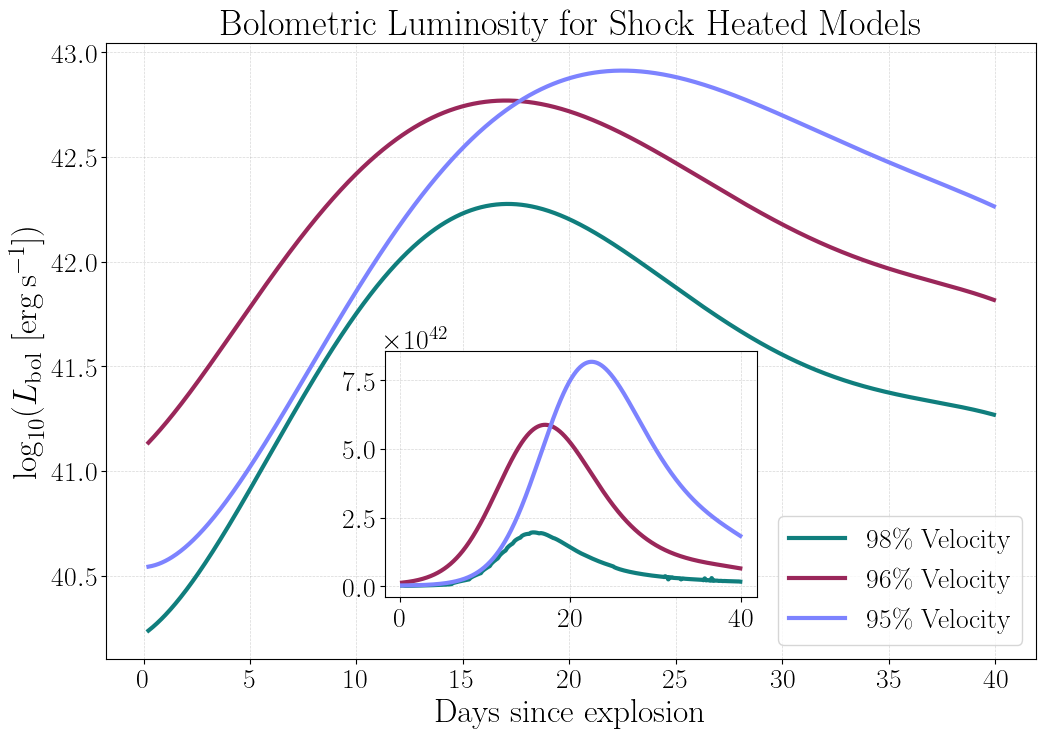}
 \caption{Bolometric light-curves heavily for shock powered models. The inset plot shows the same data without the logarithmic y-axis. These models assume that shock-heating begins as soon as the shock breaks out of the star.}
 \label{fig:shock_lc}
\end{figure}

We perform two simulations from 4h to 25d. Both simulations have mass and velocity multipliers of 1, no nickel, and unmodified inner ejecta masses. These two models are constrained to 96\% of 98\% of their original velocities. They have similar peak locations and shapes but the 96\% model peaks higher because more of its original velocity goes into shock-heating than in the 98\% model. Our final shock-heating model undergoes heating from 0 seconds to $3\times10^6$\,s ($\sim 35$\,d) and has a lower velocity limit of 95\% of its original value. These parameters create a later, higher peak than either of the other simulations. In our shock-heating runs, small numerical errors create non-physical spikes and bumps in our light-curves. To remedy the issue, we apply a univariate spline to our data without altering its original shape. These three runs are shown in Figure \ref{fig:shock_lc}.

The clumpy-medium shock-heated models have bolometric light-curves that look very similar to that of our $^{56}$Ni-powered solutions. With this model, we make assumptions on the peak of this shock-heating. Although we would expect this to be highest at earlier times when the density of the matter ejected from the star is highest, this really depends on the details of the stellar mass ejections/winds. The shock-heating will also depend on the velocity profile of the ejecta.


\subsection{Multiple Sources}

\begin{figure}
 \centering
 \includegraphics[width=\linewidth]{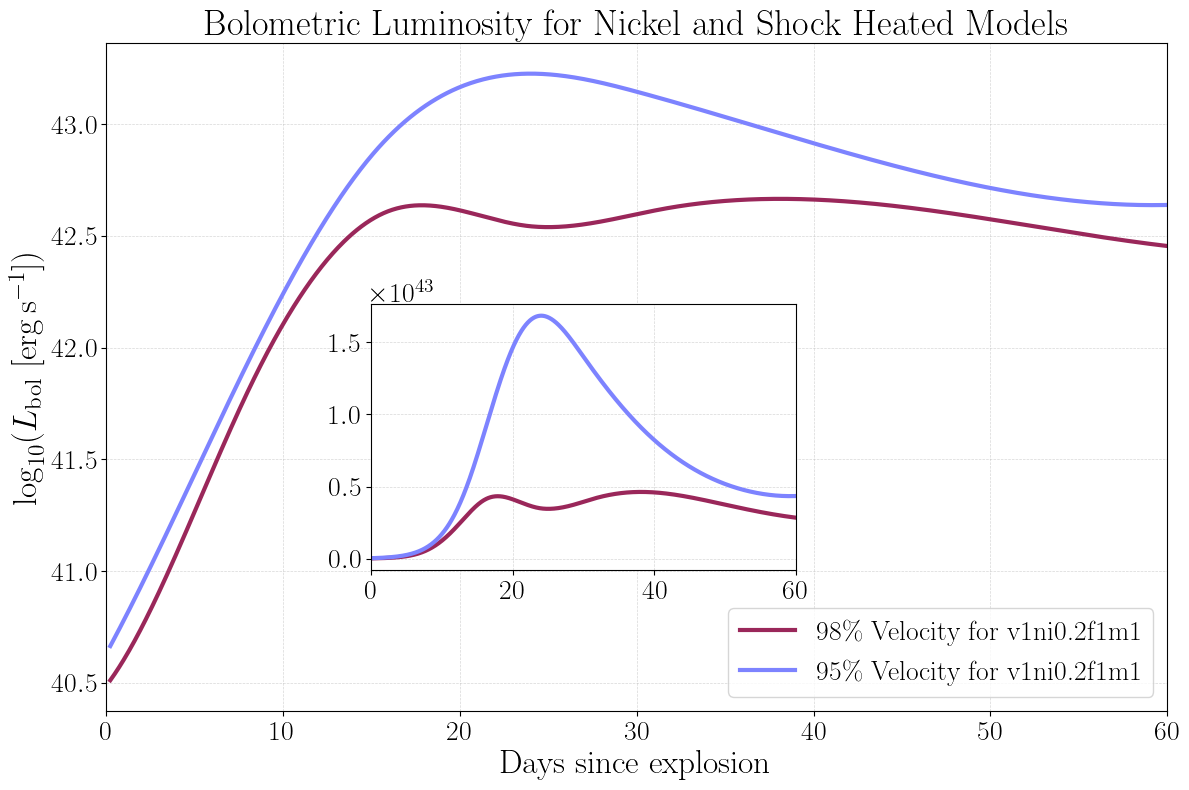}
 \caption{Nickel and shock heated runs with different shock-heating durations and powers. The inset plot shows the same data without the logarithmic y-axis.}
 \label{fig:nishock}
\end{figure}

We include a handful of runs that powered by both shock-heating and $^{56}$Ni-decay energy sources. In these runs, shock interactions determine the behavior of the light-curve at early times, while nickel-heating dominates at late times. For runs with mass multiplier 1, nickel based energy takes long enough to reach the photosphere of the supernova that we observe two distinct peaks for our early time shock-heating runs. For low mass models, the $^{56}$Ni power produces the second peak. In high mass models,, nickel-heating makes the tail of the shock light-curve broader and completely dominates at late times. Examples of tail broadened and double peaked light-curves are shown in Figure \ref{fig:nishock}.

Finally, we modeled a series of runs with the magnetar engine.  In all of these models, we assumed the magnetar turned on at shock breakout and allowed the energy source to decay with the spin down of the neutron star (we did not include the possible collapse to a black hole).  We varied both the spin rate and the magnetic field.  The luminosity of all of these models continued to rise through the end of our 60\,d simulations.  As such power sources are not ideal for Ic-BL SNe, we do not consider them further in this paper.

\subsection{Physics effects:  Opacities, Energy Deposition, Shock Heating}

\begin{figure}
 \centering
 \includegraphics[width=\linewidth]{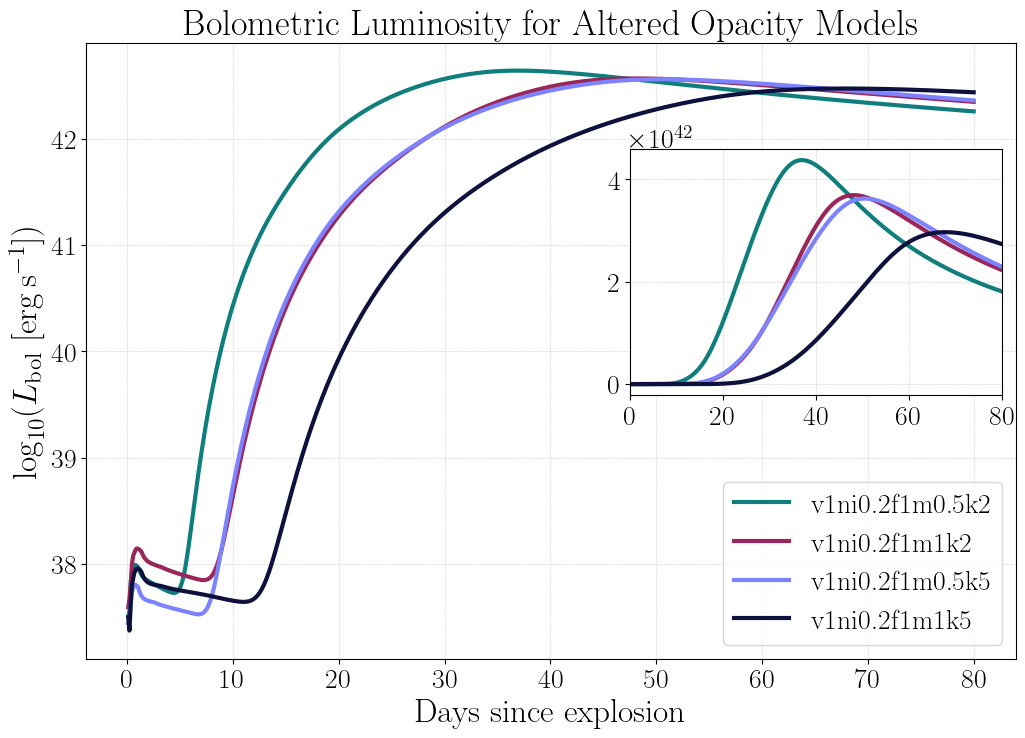}
 \caption{Light-curves from 4 models where the opacity is increased by a factor of 2 and 5 for our standard and half-mass ejecta models (m1, m0.5).  The light-curve from the m0.5 model with 5 times the opacity is nearly identical to the m1 model with 2 times the opacity.}
 \label{fig:kappa}
\end{figure}

As we have discussed in Section~\ref{sec:transport}, in our models we assume a simple 50-50, C/O, composition with line-binned opacities.  Adding a full set of elements and/or implementing the opacities using expansion-opacity methods can alter the opacity used in our transport calculations.  To test the importance of the opacities, we have artificially increased the opacity of our models by a factor of 2 and 5 from our line-binned opacities (Figure~\ref{fig:kappa}).  Under the Arnett law~\citep{1982ApJ...253..785A}, there is a degeneracy in the shape of the light-curve between ejecta mass and opacity:  the light-curve shape is proportional to $\kappa M_{\rm ejecta}/v_{\rm ejecta}$ where $\kappa$ is the opacity, $M_{\rm ejecta}$ is the ejecta mass and $v_{\rm ejecta}$ is the ejecta velocity.  This demonstrates that accurate opacities (and hence compositions) are needed to infer exact masses from supernova explosions.  In this paper, we focused on trends and, although the opacity will affect the results, it will not affect the trends discussed in the paper.  The higher opacities expected from expansion-opacity methods~\citep{2017hsn..book..843B} exacerbate the problem with Ni-only explosions to match the short rise/fall timescales of observed supernovae (see Section~\ref{sec:comparison}).

\begin{figure}
 \centering
 \includegraphics[width=\linewidth]{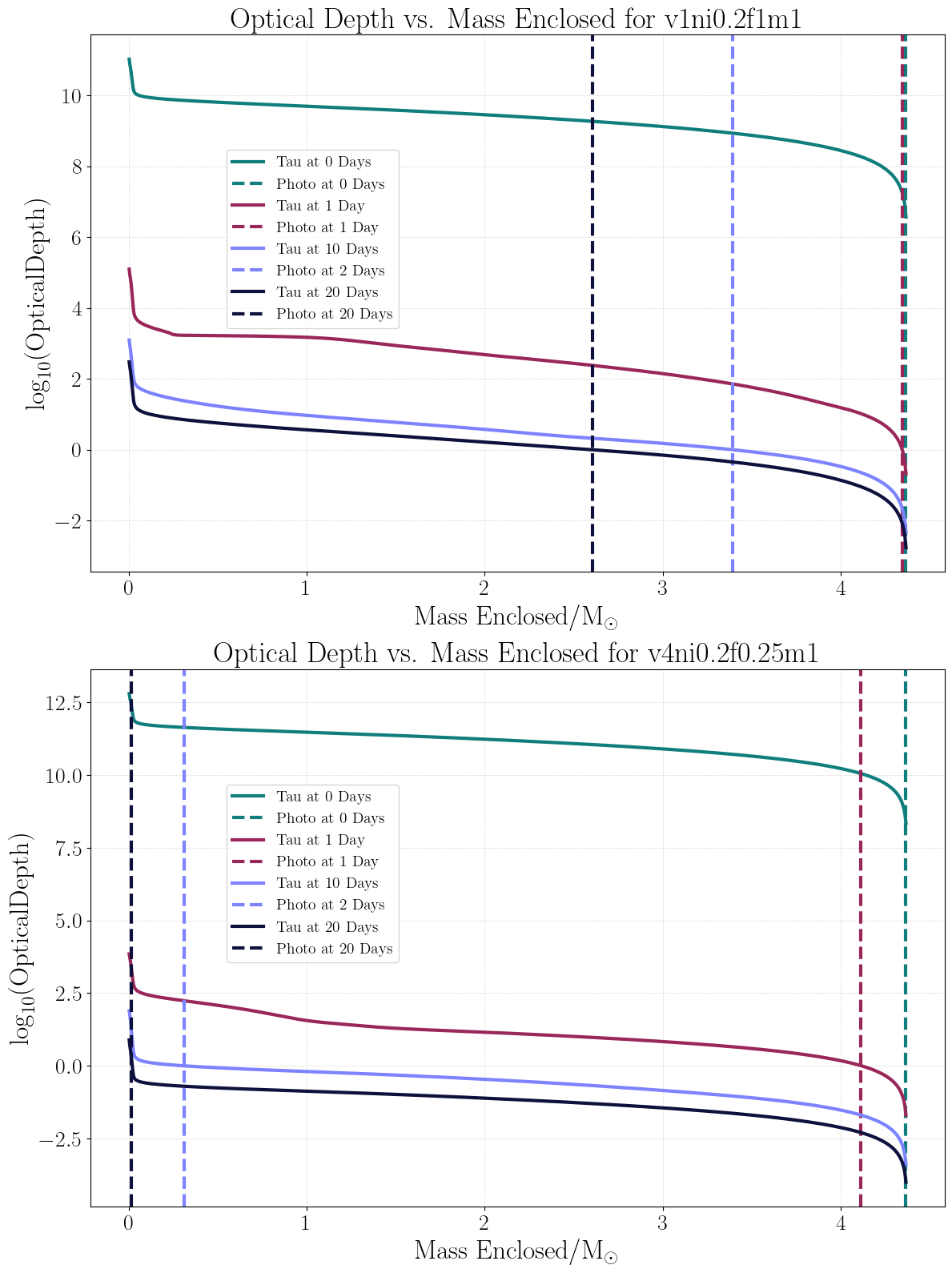}
 \caption{ Optical depth of the ejecta as a function of enclosed mass at a series of times from 0\,d to 20\,d.  As the star expands, the photosphere moves inward.  Beyond the photosphere, many of the assumptions in both our radioactive decay and shock heating models begin to fail.  For the radioactive decay model, we underestimate the heating at low optical depths leading to light-curves that will fall more slowly than our predicted models.  For our shock heating models, equilibrium between the electrons and photons will not be achieved and we will under-estimate the ultraviolet flux from these models.}
 \label{fig:tau}
\end{figure}

Our simplistic Ni-decay deposition model begins to underestimate the heating as the region of low optical-depth material increases.  Figure~\ref{fig:tau} shows the optical depth of the ejecta of two supernova models for a series of time snapshots. As the ejecta becomes optically thin in ejecta regions containing $^{56}$Ni, we will begin to slightly underestimate the total Ni-heating.  Our models will hence decay faster than what we'd expect from more realistic models.  As with our opacity approximation, improvements in our Ni-heating will produce more late-time heating, delaying the fall time of our light-curves making it more difficult for Ni-only models to fit the data (see Section~\ref{sec:comparison}).

As the optical depth of the ejecta in our models decreases, the assumption that the electrons and photons are in temperature equilibrium is no longer true.  Although the assumed energy converted from shock deceleration is the same, it will likely be converted to higher-temperature photons - e.g. in the ultraviolet.

\section{Comparison to Observation}

\begin{figure}
 \centering
 \includegraphics[width=\linewidth]{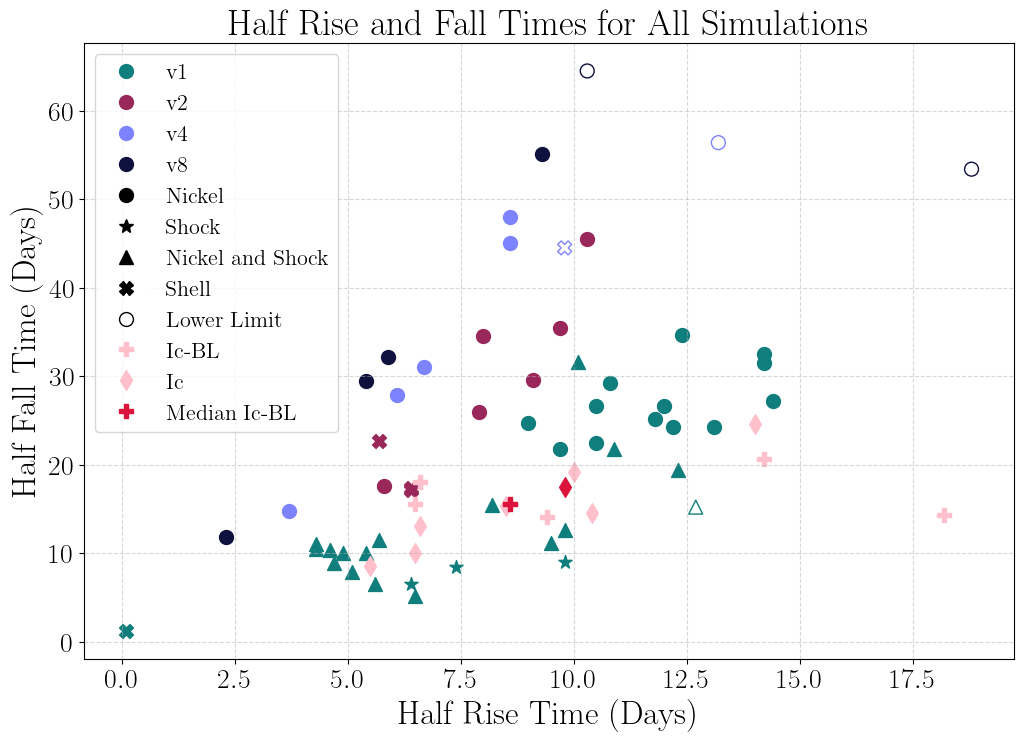}
 \caption{Rise and fall times for all nickel and shock-powered simulations. Runs with no heating source are omitted. The Ic-BL median is taken from \citet{10.1093/mnras/sty3399}. Ic-BL and Ic data points are derived from BVRI pseudo-bolometric light-curves in \citet{10.1093/mnras/stw299}.}
 \label{fig:rise_times}
\end{figure}

One way to test our models is to compare them to the current database of observed Ib/c light-curves. According to a survey of type Ib/c supernovae in \cite{10.1093/mnras/sty3399}, the median time for Ic-BL SNe to go from half luminosity (t$_{-1/2}$) to peak luminosity is 8.6 days, while the median time to return to half luminosity post peak (t$_{+1/2}$) is 15.5 days. For typical type Ic SNe, \cite{10.1093/mnras/sty3399} find t$_{-1/2}$ and t$_{+1/2}$ to be 9.8 and 17.5 days respectively. We used these values as a starting point to assess the validity of our models. Rise and fall times are given for nickel-powered runs and shock-powered runs in Tables \ref{table_nickel} and \ref{table_shock}, respectively. These results are shown graphically in Figure \ref{fig:rise_times}.

A general trend in our results is that models that include shock-heating tend to have faster decay times than those solely powered by $^{56}$Ni decay. Although these shock-heated models are better fits to the data rise and fall times from the Ic data~\citep{10.1093/mnras/sty3399}, it is certainly possible to tune the $^{56}$Ni and mass distributions to fit the data with $^{56}$Ni-heating alone. 

\begin{figure}
 \centering
 \includegraphics[width=\linewidth]{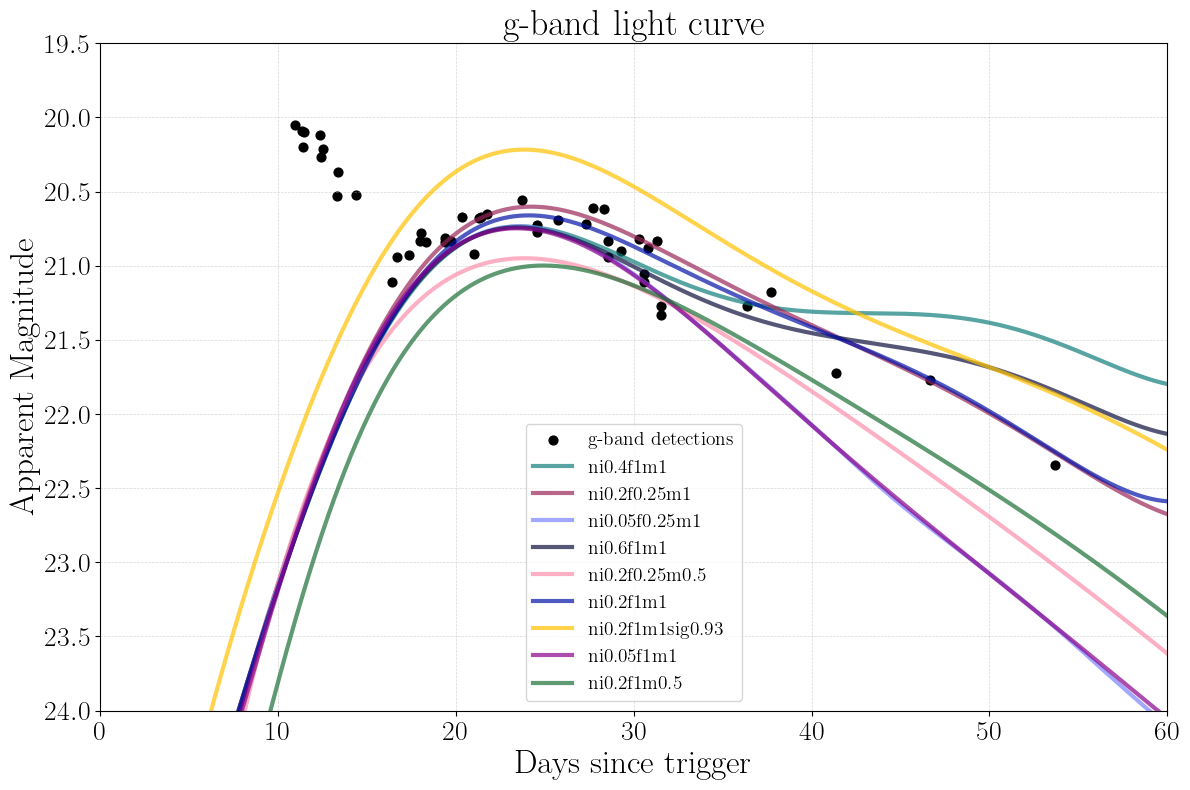}
 \caption{ g-band light-curves from a set of modeled light-curves powered by both shocks and Nickel decay.  Our default model assumes the post-shock velocities are reduced to 95\% of the original velocity.  We vary the total ejecta mass, the $^{56}$Ni mass and the $^{56}$Ni distribution.  The amount of shock heating is tuned to fit the peak of the light-curve.  Although our best fit combines our standard shock heating model reducing the velocity by 95\% and converting this kinetic energy into thermal energy and 0.2\,M$_\odot$ of $^{56}$Ni.  Combined, these sources fit all but the initial prompt emission in the light curve.}
 \label{fig:lc2025kg}
\end{figure}

With the simplifications in our calculations, we have focused on studying trends and rough comparisons to data.  However, it is worth comparing directly to a supernova light-curve.  Figure~\ref{fig:lc2025kg} shows a series of models fit to SN 2025kg~\citep{2025ApJ...988L..13R}.  This supernova is believed to be a jet-driven supernova and the initial emergence of the shock~\citep{2025ApJ...988L..14E}.  Our models focus on fitting the rise and fall of the peak light-curve.  Our default model assumes the post-shock velocities are reduced to 95\% of the original velocity.  We then vary the the total ejecta mass, the nickel mass and its distribution.  The peak is dominated by the shock heating and the nickel shapes the fall off of the light-curve.  In our models, too much nickel causes the light curve to be too bright after 40\,d and too little makes the light curve fall off too quickly. Our best fit includes 0.2\,M$_\odot$ of $^{56}$Ni, at the low end of the predictions by the models used in \cite{2025ApJ...988L..13R}.  If we increase the shock heating too much, our light-curve is too bright.

\subsection{Distinguishing Sources}
\label{sec:comparison}

As we have discussed above, supernova light-curves are affected by a wide range of explosion properties:  energy sources (radioactive decay, long-lived central engines, shocks), energetics and mass of the ejecta, ejecta profiles (both the mixing of material and the density/temperature profile of the star set by the explosion and the stellar progenitor).  If we observe only the bolometric luminosity, it would be difficult to disentangle many of these explosion properties.  The Arnett law argues that, under certain approximate solutions, the shape of the light curve has existing degeneracies between the opacity, ejecta mass and ejecta velocity.  But, by leveraging the broad set of data on supernovae, we can break some of these degeneracies.  For instance, we typically have a wide energy range from infra-red to ultraviolet.  Although early-time X-rays are less common, they can also place constraints on the explosion and its progenitor.  In addition, spectral features can probe both the temperature and, through line profiles, the velocity of the ejecta.  We discuss this data and its constraints on the observations here.

From the bolometric light-curves alone, it is difficult to distinguish between different energy sources. Here we look at some observations that may differentiate these sources. For our shock-heating models where we assume clumpy winds that immediately add energy to the blast wave, the early time light-curve will be much brighter than that of a solely nickel-powered run (which has to wait for the $^{56}$Ni energy to diffuse out to the photosphere). These early time differences are easily seen in comparing the bolometric light-curves from Figure~\ref{fig:ni_lc} to those in Figure~\ref{fig:shock_lc}. However, we do have models where the shock-heating is delayed (Figure~\ref{fig:shellshock}) or the $^{56}$Ni is extensively mixed outward ($f0.1$ models) that defy these trends.

Alternatively, late-time observations could help us distinguish between the two models. In our models, $^{56}$Ni-decay dominates the late-time emission. $^{56}$Ni decay will continue to deposit energy both through deposition in the innermost ejecta and through the energy injected by positrons (which are likely to remain trapped in the ejecta at late-times, depositing their energy to this ejecta). Here, the velocity of the innermost ejecta can drastically alter the light-curves. A major caveat using late-time observations to compare results is the fact that, currently, supernova light-curve codes can get very different results at late times~\citep{2022A&A...668A.163B} and these numerical artifacts must be understood before we can do detailed late-time analyses to study ejecta properties.

\begin{figure}
 \centering
 \includegraphics[width=\linewidth]{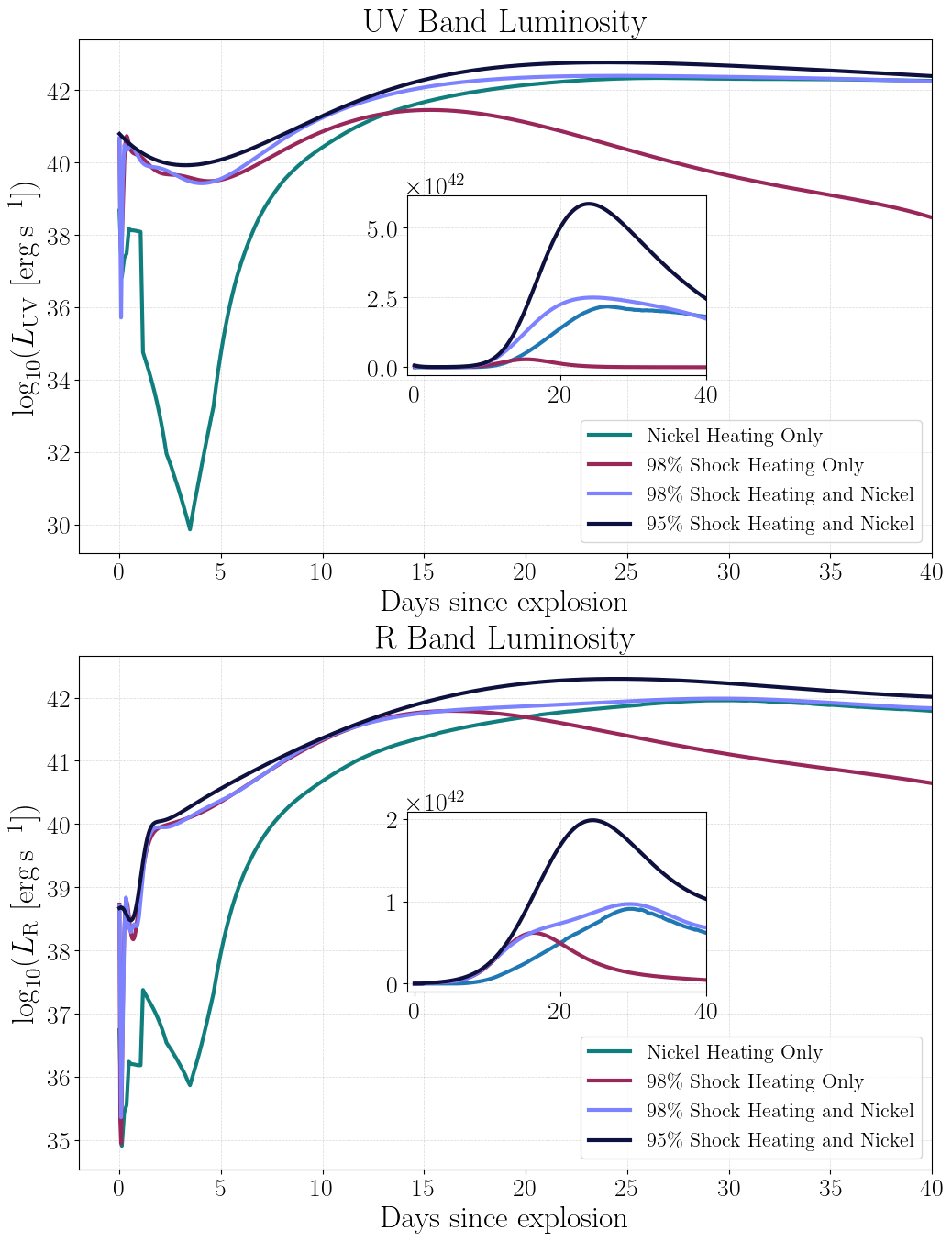}
 \caption{R-band and Swift's UVOT-band luminosities for v1ni0.2f1m0.5, v1ni0m0.5 with 98\% velocity shock-heating, v1ni0.2f1m0.5 with 98\% velocity shock-heating, and v1ni0.2f1m0.5 with 98\% velocity shock-heating. The inset plot shows the same data without the logarithmic y-axis.}
 \label{fig:bands}
\end{figure}

Because the photospheric temperatures evolve differently, it is possible to use color bands to distinguish between sources. In Figure~\ref{fig:bands}, we present Swift UV and r-band luminosities for two nickel-shock runs, one nickel only run and one shock only run. Both nickel-heating and shock-heating are Swift UV bright at peak. For events of similar luminosity, distinguishing between shock-heating and nickel-heating at peak is virtually impossible. But, at early times, shock-heated systems may be much brighter than those heated by radioactive decay.  In addition, the equilibrium assumptions in our shock-heating model will, if anything, underestimate the ultraviolet flux so we expect the shock-heated models to be even higher temperatures at early times.

\begin{figure}
 \centering
 \includegraphics[width=\linewidth]{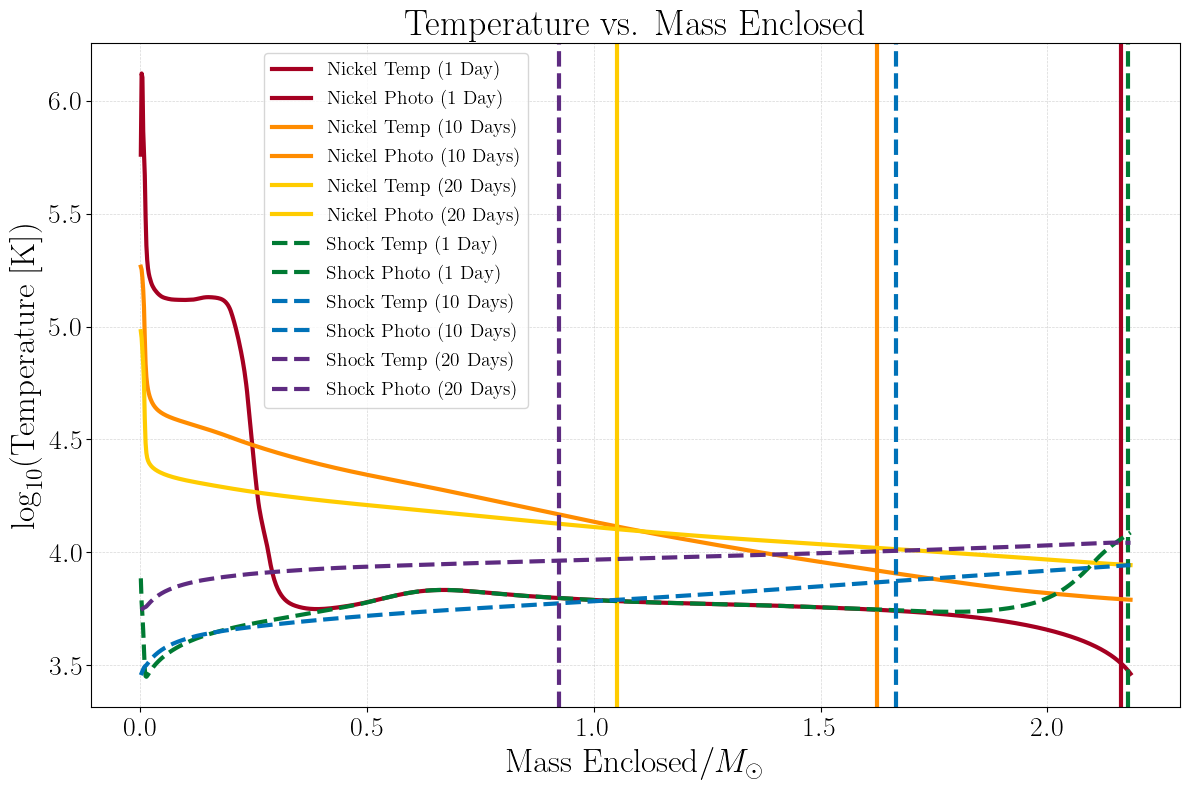}
 \caption{Temperature profile comparison for a shock-heated model and a nickel-heated model at 1, 20, and 40 days. The full parameters for the nickel and shock-heated models shown are v1ni0.2f1m0.5 and 95\% velocity with v1ni0.2f1m0.5, respectively.}
 \label{fig:shock_temp}
\end{figure}

\begin{figure}
 \centering
 \includegraphics[width=\linewidth]{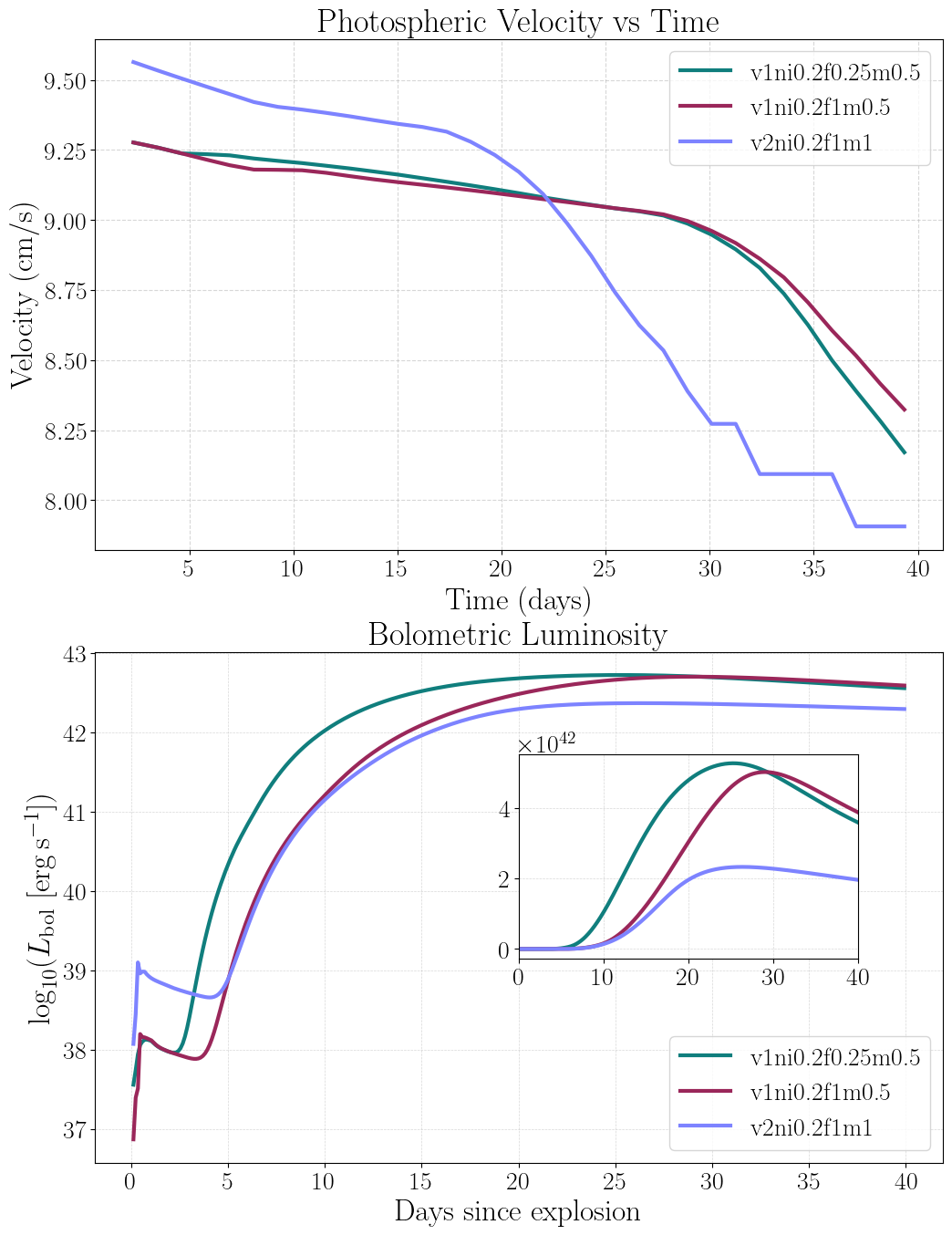}
 \caption{Luminosity versus time (bottom) and photospheric velocity (top) versus time for three models:  v1ni0.2f0.25m0.5, v1ni0.2f1m0.5, v2ni0.2f1m1.  Although the photospheric velocity can help disentangle the degeneracy in the mass and velocity (compare v1ni0.2f1m0.5 and v2ni0.2f1m1).  But it is more difficult to disentangle the amount of mixing (compare v1ni0.2f0.25m0.5 and v1ni0.2f1m0.5).  $\gamma-$ray observations or late-time iron lines could determine this mixing, but these observations are typically limited to nearby supernovae.}
 \label{fig:velcomp1}
\end{figure}

\begin{figure}
 \centering
 \includegraphics[width=\linewidth]{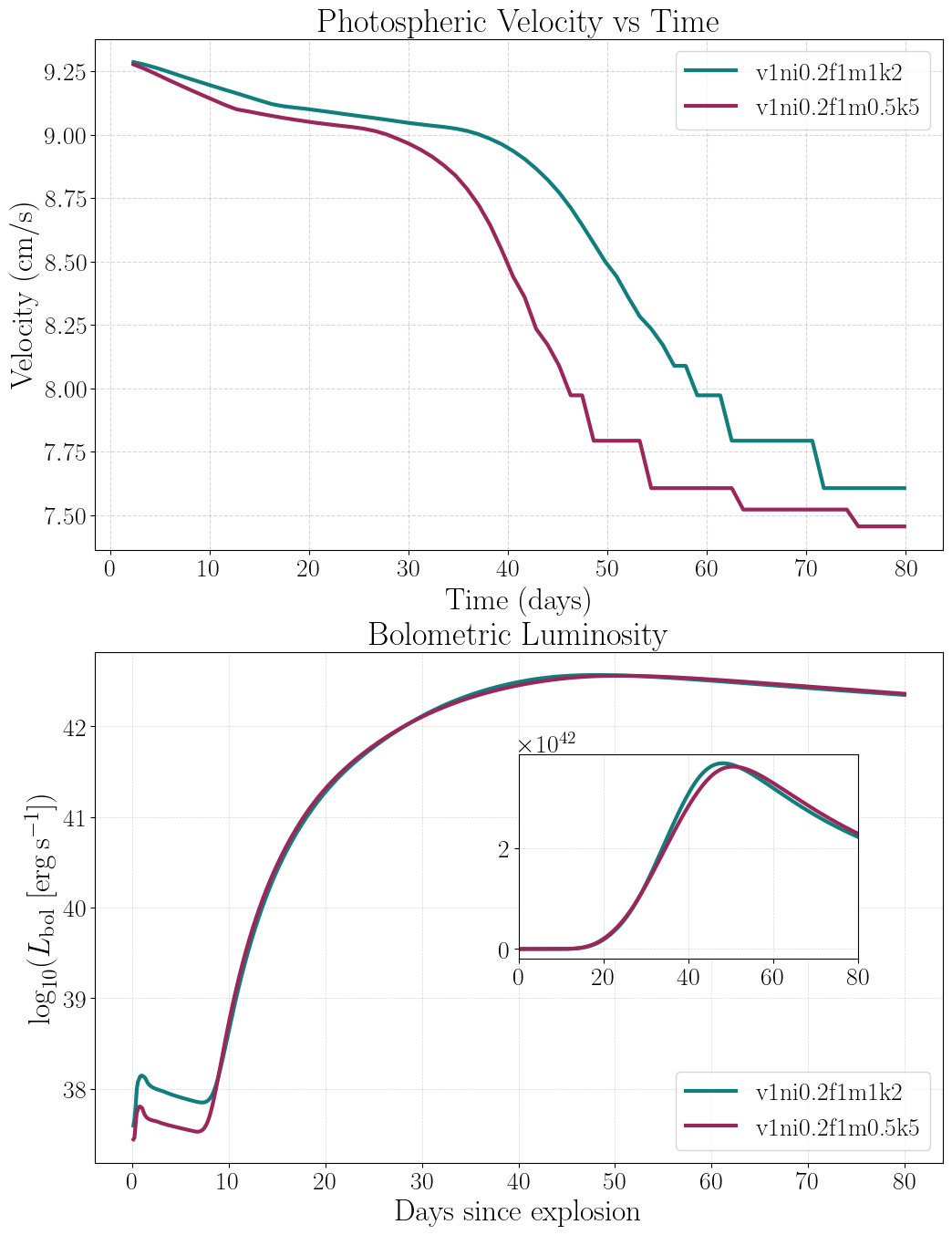}
 \caption{Luminosity versus time (bottom) and photospheric velocity (top) versus time for two of our opacity models:  v1ni0.2f1m1k2, v1ni0.2f1m0.5k5.  Although these two models have very similar light-curves, the velocity profiles are very different.}
 \label{fig:velcomp2}
\end{figure}

We defer detailed spectra calculations to a later paper and these spectra may allow us to better differentiate the models.  But we can gauge ability of spectra and, in particular, the constraints line profiles place on the ejecta velocity, to distinguish different models by comparing the photospheric velocities of models with similar light-curve properties.  Figure~\ref{fig:velcomp1} shows the luminosity and photopsheric velocity versus time for 3 different models where we have varied the mixing of $^{56}$Ni, the mass and the velocity of the ejecta.  The degeneracy in the light-curve between mass and velocity, and hence spectra, can be seen by comparing models v1ni0.2f1m0.5 and v2ni0.2f1m1.  But the photospheric velocity is less capable of distinguishing the amount of outward mixing of the $^{56}$Ni (compare models v1ni0.2f0.25m0.5 and v1ni0.2f1m0.5), in agreement with the results from~\cite{2023ApJ...956...19F}.  But measurements of the photospheric velocity can help us disentangle degeneracies in the ejecta mass and velocity.  Figure~\ref{fig:velcomp2} shows that line profiles can also be used to determine the opacity.

\section{Conclusions}

In this paper, we performed 1-dimensional nickel and shock powered Ic supernovae simulations across a wide range of masses, velocities, and engine parameters. We find that it is difficult to produce realistic light-curves using the decay of $^{56}$Ni alone. Increasing the velocity of the ejecta or lowering its mass can shorten the rise time, but no parameter adjustments in our models are able to adequately shorten the fall time of the light-curve. Tuning the velocity distribution of the matter, or more importantly, the $^{56}$Ni distribution, may be able to match the rapid fall time (e.g. placing the $^{56}$Ni primarily in the outer ejecta). 

Given this result, we explore shock-heating as an alternative. By adjusting the period of shock-heating and the percentage of velocity converted to kinetic energy, we are able to fit a wide variety of different light-curves and adjust our rise and falls times. With a combination of shock-heating and nickel-power, we use shock-heating to power the early light-curve and allow nickel-heating to broaden the tail, achieving a wide range of peak luminosities and light-curve shapes. To observationally distinguish between nickel-heating and shock-heating, we recommend early (shock-heating dominated) and late (nickel-heating dominated) observations. 

Central engine power sources (magnetar, fallback) also dominate primarily the late-time light-curve. Their energy must diffuse through the star to power the observed emission. It is unlikely these mechanisms can power the light-curves in our comparison sample unless the ejecta mass is extremely low and the innermost ejecta velocity is high.

Future work can focus on expanding the parameter space for shock-heating simulations. By altering the mechanism used to lower the velocity (Equation \ref{eq-sig}, in this case), a wide variety of different light-curves can be fit using this method. Additionally, changing the time period over which shock-heating takes place can flatten and raise the peak. Exploring different heating mechanisms and distinguishing between them observationally needs to be done in greater detail. 

More importantly, detailed studies of the ejecta velocity distribution are crucial in determining the breadth of solutions that a given power source can produce. Future studies will focus on these velocity distributions to better understand the range of possible light-curves. Observations can then be used to constrain the ejecta properties based on the power-source, providing new insight into the explosion mechanism.

In addition, Ic-BL SNe are almost certainly highly asymmetric and multi-dimensional models are needed to truly compare to observations.
Finally, we defer detailed work on the spectra (including out of equilibrium effects) to later work.

\begin{acknowledgements}

The work by AEN and CLF was supported by the US Department of Energy through the Los Alamos National Laboratory. Los Alamos National Laboratory is operated by Triad National Security, LLC, for the National Nuclear Security Administration of U.S.\ Department of Energy (Contract No.\ 89233218CNA000001).  We thank Jillian Rastinejad for providing us the light curve data from SN 2025kg.

\end{acknowledgements}

\bibliography{refs}{}

\begin{thebibliography}{}
\expandafter\ifx\csname natexlab\endcsname\relax\def\natexlab#1{#1}\fi
\providecommand{\url}[1]{\href{#1}{#1}}
\providecommand{\dodoi}[1]{doi:~\href{http://doi.org/#1}{\nolinkurl{#1}}}
\providecommand{\doeprint}[1]{\href{http://ascl.net/#1}{\nolinkurl{http://ascl.net/#1}}}
\providecommand{\doarXiv}[1]{\href{https://arxiv.org/abs/#1}{\nolinkurl{https://arxiv.org/abs/#1}}}

\bibitem[{Anand {et~al.}(2024)Anand, Barnes, Yang, Kasliwal, Coughlin,
  Sollerman, De, Fremling, Corsi, Ho, Balasubramanian, Omand, Srinivasaragavan,
  Cenko, Ahumada, Andreoni, Dahiwale, Das, Jencson, Karambelkar, Kumar,
  Metzger, Perley, Sarin, Schweyer, Schulze, Sharma, Sit, Stein, Tartaglia,
  Tinyanont, Tzanidakis, van Roestel, Yao, Bloom, Cook, Dekany, Graham, Groom,
  Kaplan, Masci, Medford, Riddle, \& Zhang}]{Anand_2024}
Anand, S., Barnes, J., Yang, S., {et~al.} 2024, The Astrophysical Journal, 962,
  68, \dodoi{10.3847/1538-4357/ad11df}

\bibitem[{{Arnett}(1982)}]{1982ApJ...253..785A}
{Arnett}, W.~D. 1982, \apj, 253, 785, \dodoi{10.1086/159681}

\bibitem[{{Arnett} {et~al.}(2017){Arnett}, {Fryer}, \&
  {Matheson}}]{2017ApJ...846...33A}
{Arnett}, W.~D., {Fryer}, C., \& {Matheson}, T. 2017, \apj, 846, 33,
  \dodoi{10.3847/1538-4357/aa8173}

\bibitem[{{Blinnikov}(2017)}]{2017hsn..book..843B}
{Blinnikov}, S. 2017, in Handbook of Supernovae, ed. A.~W. {Alsabti} \&
  P.~{Murdin}, 843, \dodoi{10.1007/978-3-319-21846-5_31}

\bibitem[{{Blinnikov} \& {Bartunov}(2011)}]{2011ascl.soft08013B}
{Blinnikov}, S.~I., \& {Bartunov}, O.~S. 2011, {STELLA: Multi-group Radiation
  Hydrodynamics Code}, Astrophysics Source Code Library, record ascl:1108.013

\bibitem[{{Blondin} {et~al.}(2022){Blondin}, {Blinnikov}, {Callan}, {Collins},
  {Dessart}, {Even}, {Fl{\"o}rs}, {Fullard}, {Hillier}, {Jerkstrand}, {Kasen},
  {Katz}, {Kerzendorf}, {Kozyreva}, {O'Brien}, {P{\'a}ssaro}, {Roth}, {Shen},
  {Shingles}, {Sim}, {Singhal}, {Smith}, {Sorokina}, {Utrobin}, {Vogl},
  {Williamson}, {Wollaeger}, {Woosley}, \& {Wygoda}}]{2022A&A...668A.163B}
{Blondin}, S., {Blinnikov}, S., {Callan}, F.~P., {et~al.} 2022, \aap, 668,
  A163, \dodoi{10.1051/0004-6361/202244134}

\bibitem[{{Brown} {et~al.}(2023){Brown}, {Robertson}, {Devarakonda}, {Sarria},
  {Pooley}, \& {Stritzinger}}]{2023Univ....9..218B}
{Brown}, P.~J., {Robertson}, M., {Devarakonda}, Y., {et~al.} 2023, Universe, 9,
  218, \dodoi{10.3390/universe9050218}

\bibitem[{{Cano} {et~al.}(2017){Cano}, {Wang}, {Dai}, \& {Wu}}]{Cano2017}
{Cano}, Z., {Wang}, S.-Q., {Dai}, Z.-G., \& {Wu}, X.-F. 2017, Advances in
  Astronomy, 2017, 8929054, \dodoi{10.1155/2017/8929054}

\bibitem[{{Castor}(2004)}]{2004rahy.book.....C}
{Castor}, J.~I. 2004, {Radiation Hydrodynamics}

\bibitem[{{Chatzopoulos} \& {Tuminello}(2019)}]{2019ApJ...874...68C}
{Chatzopoulos}, E., \& {Tuminello}, R. 2019, \apj, 874, 68,
  \dodoi{10.3847/1538-4357/ab0ae6}

\bibitem[{{Chatzopoulos} \& {Weide}(2019)}]{2019ApJ...876..148C}
{Chatzopoulos}, E., \& {Weide}, K. 2019, \apj, 876, 148,
  \dodoi{10.3847/1538-4357/ab18f9}

\bibitem[{{Chevalier}(1989)}]{1989ApJ...346..847C}
{Chevalier}, R.~A. 1989, \apj, 346, 847, \dodoi{10.1086/168066}

\bibitem[{{Chiba} \& {Moriya}(2024)}]{2024ApJ...973...14C}
{Chiba}, R., \& {Moriya}, T.~J. 2024, \apj, 973, 14,
  \dodoi{10.3847/1538-4357/ad6c37}

\bibitem[{{Colgan} {et~al.}(2016){Colgan}, {Kilcrease}, {Magee}, {Sherrill},
  {Abdallah}, {Hakel}, {Fontes}, {Guzik}, \& {Mussack}}]{2016ApJ...817..116C}
{Colgan}, J., {Kilcrease}, D.~P., {Magee}, N.~H., {et~al.} 2016, \apj, 817,
  116, \dodoi{10.3847/0004-637X/817/2/116}

\bibitem[{{De La Rosa} {et~al.}(2017){De La Rosa}, {Roming}, \&
  {Fryer}}]{2017ApJ...850..133D}
{De La Rosa}, J., {Roming}, P., \& {Fryer}, C. 2017, \apj, 850, 133,
  \dodoi{10.3847/1538-4357/aa93ee}

\bibitem[{{Dexter} \& {Kasen}(2013)}]{2013ApJ...772...30D}
{Dexter}, J., \& {Kasen}, D. 2013, \apj, 772, 30,
  \dodoi{10.1088/0004-637X/772/1/30}

\bibitem[{{Eyles-Ferris} {et~al.}(2025){Eyles-Ferris}, {Jonker}, {Levan},
  {Malesani}, {Sarin}, {Fryer}, {Rastinejad}, {Burns}, {Tanvir}, {O'Brien},
  {Fong}, {Mandel}, {Gompertz}, {Kilpatrick}, {Bloemen}, {Bright},
  {Carotenuto}, {Corcoran}, {Cotter}, {Groot}, {Izzo}, {Laskar},
  {Martin-Carrillo}, {Palmerio}, {Ravasio}, {van Roestel}, {Saccardi},
  {Starling}, {Thakur}, {Vergani}, {Vreeswijk}, {Bauer}, {Campana},
  {Chac{\'o}n}, {Chrimes}, {Covino}, {van Dalen}, {D'Elia}, {De Pasquale},
  {Habeeb}, {Hartmann}, {van Hoof}, {Jakobsson}, {Julakanti}, {Leloudas}, {Mata
  S{\'a}nchez}, {Nixon}, {Pieterse}, {Pugliese}, {Quirola-V{\'a}squez},
  {Rayson}, {Salvaterra}, {Schneider}, {Torres}, \&
  {Zafar}}]{2025ApJ...988L..14E}
{Eyles-Ferris}, R. A.~J., {Jonker}, P.~G., {Levan}, A.~J., {et~al.} 2025,
  \apjl, 988, L14, \dodoi{10.3847/2041-8213/ade1d9}

\bibitem[{{Fontes} {et~al.}(2020){Fontes}, {Fryer}, {Hungerford}, {Wollaeger},
  \& {Korobkin}}]{2020MNRAS.493.4143F}
{Fontes}, C.~J., {Fryer}, C.~L., {Hungerford}, A.~L., {Wollaeger}, R.~T., \&
  {Korobkin}, O. 2020, \mnras, 493, 4143, \dodoi{10.1093/mnras/staa485}

\bibitem[{{Frey} {et~al.}(2013){Frey}, {Even}, {Whalen}, {Fryer}, {Hungerford},
  {Fontes}, \& {Colgan}}]{2013ApJS..204...16F}
{Frey}, L.~H., {Even}, W., {Whalen}, D.~J., {et~al.} 2013, \apjs, 204, 16,
  \dodoi{10.1088/0067-0049/204/2/16}

\bibitem[{Fryer \& Fryer(2024)}]{fryer224}
Fryer, C., \& Fryer, D.~A. 2024, International Journal of Modern Physics D, 0,
  null, \dodoi{10.1142/S0218271825400024}

\bibitem[{{Fryer}(1999)}]{1999ApJ...522..413F}
{Fryer}, C.~L. 1999, \apj, 522, 413, \dodoi{10.1086/307647}

\bibitem[{{Fryer}(2009)}]{2009ApJ...699..409F}
---. 2009, \apj, 699, 409, \dodoi{10.1088/0004-637X/699/1/409}

\bibitem[{{Fryer} {et~al.}(2018){Fryer}, {Andrews}, {Even}, {Heger}, \&
  {Safi-Harb}}]{2018ApJ...856...63F}
{Fryer}, C.~L., {Andrews}, S., {Even}, W., {Heger}, A., \& {Safi-Harb}, S.
  2018, \apj, 856, 63, \dodoi{10.3847/1538-4357/aaaf6f}

\bibitem[{Fryer {et~al.}(2024)Fryer, Burns, Ho, Corsi, Lien, Perley, Vail, \&
  Villar}]{fryer2024explainingnonmergergammaraybursts}
Fryer, C.~L., Burns, E., Ho, A. Y.~Q., {et~al.} 2024, Explaining Non-Merger
  Gamma-Ray Bursts and Broad-Lined Supernovae with Close Binary Progenitors
  with Black Hole Central Engine.
\newblock \doarXiv{2410.10378}

\bibitem[{{Fryer} {et~al.}(1999){Fryer}, {Colgate}, \&
  {Pinto}}]{1999ApJ...511..885F}
{Fryer}, C.~L., {Colgate}, S.~A., \& {Pinto}, P.~A. 1999, \apj, 511, 885,
  \dodoi{10.1086/306701}

\bibitem[{Fryer {et~al.}(2020)Fryer, Fontes, Warsa, Roming, Coffing, \&
  Wood}]{Fryer_2020}
Fryer, C.~L., Fontes, C.~J., Warsa, J.~S., {et~al.} 2020, The Astrophysical
  Journal, 898, 123, \dodoi{10.3847/1538-4357/ab99a7}

\bibitem[{{Fryer} {et~al.}(2019){Fryer}, {Lloyd-Ronning}, {Wollaeger},
  {Wiggins}, {Miller}, {Dolence}, {Ryan}, \& {Fields}}]{2019EPJA...55..132F}
{Fryer}, C.~L., {Lloyd-Ronning}, N., {Wollaeger}, R., {et~al.} 2019, European
  Physical Journal A, 55, 132, \dodoi{10.1140/epja/i2019-12818-y}

\bibitem[{{Fryer} {et~al.}(2006){Fryer}, {Young}, \&
  {Hungerford}}]{2006ApJ...650.1028F}
{Fryer}, C.~L., {Young}, P.~A., \& {Hungerford}, A.~L. 2006, \apj, 650, 1028,
  \dodoi{10.1086/506250}

\bibitem[{{Fryer} {et~al.}(2023{\natexlab{a}}){Fryer}, {Keiter}, {Sharma},
  {Leveillee}, {Meyerhofer}, {Barnak}, {Byvank}, {Elshafiey}, {Fontes},
  {Johns}, {Kozlowski}, \& {Urbatsch}}]{2023arXiv231216677F}
{Fryer}, C.~L., {Keiter}, P.~A., {Sharma}, V., {et~al.} 2023{\natexlab{a}},
  arXiv e-prints, arXiv:2312.16677, \dodoi{10.48550/arXiv.2312.16677}

\bibitem[{{Fryer} {et~al.}(2023{\natexlab{b}}){Fryer}, {Burns}, {Hungerford},
  {Safi-Harb}, {Wollaeger}, {Miller}, {Negro}, {Anandagoda}, \&
  {Hartmann}}]{2023ApJ...956...19F}
{Fryer}, C.~L., {Burns}, E., {Hungerford}, A., {et~al.} 2023{\natexlab{b}},
  \apj, 956, 19, \dodoi{10.3847/1538-4357/ace0c3}

\bibitem[{{Fryer} {et~al.}(2024){Fryer}, {Hungerford}, {Wollaeger}, {Miller},
  {De}, {Fontes}, {Korobkin}, {Kedia}, {Ristic}, \&
  {O'Shaughnessy}}]{2024ApJ...961....9F}
{Fryer}, C.~L., {Hungerford}, A.~L., {Wollaeger}, R.~T., {et~al.} 2024, \apj,
  961, 9, \dodoi{10.3847/1538-4357/ad1036}

\bibitem[{{Goldberg} {et~al.}(2022){Goldberg}, {Jiang}, \&
  {Bildsten}}]{2022ApJ...933..164G}
{Goldberg}, J.~A., {Jiang}, Y.-F., \& {Bildsten}, L. 2022, \apj, 933, 164,
  \dodoi{10.3847/1538-4357/ac75e3}

\bibitem[{{Herant} {et~al.}(1994){Herant}, {Benz}, {Hix}, {Fryer}, \&
  {Colgate}}]{1994ApJ...435..339H}
{Herant}, M., {Benz}, W., {Hix}, W.~R., {Fryer}, C.~L., \& {Colgate}, S.~A.
  1994, \apj, 435, 339, \dodoi{10.1086/174817}

\bibitem[{{Ivanova} {et~al.}(2013){Ivanova}, {Justham}, {Chen}, {De Marco},
  {Fryer}, {Gaburov}, {Ge}, {Glebbeek}, {Han}, {Li}, {Lu}, {Marsh},
  {Podsiadlowski}, {Potter}, {Soker}, {Taam}, {Tauris}, {van den Heuvel}, \&
  {Webbink}}]{2013A&ARv..21...59I}
{Ivanova}, N., {Justham}, S., {Chen}, X., {et~al.} 2013, \aapr, 21, 59,
  \dodoi{10.1007/s00159-013-0059-2}

\bibitem[{{Jin} {et~al.}(2021){Jin}, {Yoon}, \&
  {Blinnikov}}]{2021ApJ...910...68J}
{Jin}, H., {Yoon}, S.-C., \& {Blinnikov}, S. 2021, \apj, 910, 68,
  \dodoi{10.3847/1538-4357/abe0b1}

\bibitem[{{Kasen} \& {Bildsten}(2010)}]{2010ApJ...717..245K}
{Kasen}, D., \& {Bildsten}, L. 2010, \apj, 717, 245,
  \dodoi{10.1088/0004-637X/717/1/245}

\bibitem[{{Korobkin} {et~al.}(2021){Korobkin}, {Wollaeger}, {Fryer},
  {Hungerford}, {Rosswog}, {Fontes}, {Mumpower}, {Chase}, {Even}, {Miller},
  {Misch}, \& {Lippuner}}]{2021ApJ...910..116K}
{Korobkin}, O., {Wollaeger}, R.~T., {Fryer}, C.~L., {et~al.} 2021, \apj, 910,
  116, \dodoi{10.3847/1538-4357/abe1b5}

\bibitem[{{Kuncarayakti, H.} {et~al.}(2023){Kuncarayakti, H.}, {Sollerman, J.},
  {Izzo, L.}, {Maeda, K.}, {Yang, S.}, {Schulze, S.}, {Angus, C. R.}, {Aubert,
  M.}, {Auchettl, K.}, {Della Valle, M.}, {Dessart, L.}, {Hinds, K.}, {Kankare,
  E.}, {Kawabata, M.}, {Lundqvist, P.}, {Nakaoka, T.}, {Perley, D.}, {Raimundo,
  S. I.}, {Strotjohann, N. L.}, {Taguchi, K.}, {Cai, Y.-Z.}, {Charalampopoulos,
  P.}, {Fang, Q.}, {Fraser, M.}, {Gutiérrez, C. P.}, {Imazawa, R.}, {Kangas,
  T.}, {Kawabata, K. S.}, {Kotak, R.}, {Kravtsov, T.}, {Matilainen, K.},
  {Mattila, S.}, {Moran, S.}, {Murata, I.}, {Salmaso, I.}, {Anderson, J. P.},
  {Ashall, C.}, {Bellm, E. C.}, {Benetti, S.}, {Chambers, K. C.}, {Chen,
  T.-W.}, {Coughlin, M.}, {De Colle, F.}, {Fremling, C.}, {Galbany, L.},
  {Gal-Yam, A.}, {Gromadzki, M.}, {Groom, S. L.}, {Hajela, A.}, {Inserra, C.},
  {Kasliwal, M. M.}, {Mahabal, A. A.}, {Martin-Carrillo, A.}, {Moore, T.},
  {Müller-Bravo, T. E.}, {Nicholl, M.}, {Ragosta, F.}, {Riddle, R. L.},
  {Sharma, Y.}, {Srivastav, S.}, {Stritzinger, M. D.}, {Wold, A.}, \& {Young,
  D. R.}}]{refId0}
{Kuncarayakti, H.}, {Sollerman, J.}, {Izzo, L.}, {et~al.} 2023, A\&A, 678,
  A209, \dodoi{10.1051/0004-6361/202346526}

\bibitem[{Menegazzi {et~al.}(2024)Menegazzi, Fujibayashi, Shibata, Betranhandy,
  \& Takahashi}]{menegazzi2024varietydiscwinddrivenexplosions}
Menegazzi, L.~C., Fujibayashi, S., Shibata, M., Betranhandy, A., \& Takahashi,
  K. 2024, Variety of disc wind-driven explosions in massive rotating stars.
  II. Dependence on the progenitor.
\newblock \doarXiv{2411.04221}

\bibitem[{{Metzger} \& {Piro}(2014)}]{2014MNRAS.439.3916M}
{Metzger}, B.~D., \& {Piro}, A.~L. 2014, \mnras, 439, 3916,
  \dodoi{10.1093/mnras/stu247}

\bibitem[{Modjaz {et~al.}(2008)Modjaz, Kewley, Kirshner, Stanek, Challis,
  Garnavich, Greene, Kelly, \& Prieto}]{Modjaz_2008}
Modjaz, M., Kewley, L., Kirshner, R.~P., {et~al.} 2008, The Astronomical
  Journal, 135, 1136–1150, \dodoi{10.1088/0004-6256/135/4/1136}

\bibitem[{{Moriya} {et~al.}(2019){Moriya}, {M{\"u}ller}, {Chan}, {Heger}, \&
  {Blinnikov}}]{2019ApJ...880...21M}
{Moriya}, T.~J., {M{\"u}ller}, B., {Chan}, C., {Heger}, A., \& {Blinnikov},
  S.~I. 2019, \apj, 880, 21, \dodoi{10.3847/1538-4357/ab2643}

\bibitem[{{Nakamura} {et~al.}(2001){Nakamura}, {Umeda}, {Iwamoto}, {Nomoto},
  {Hashimoto}, {Hix}, \& {Thielemann}}]{2001ApJ...555..880N}
{Nakamura}, T., {Umeda}, H., {Iwamoto}, K., {et~al.} 2001, \apj, 555, 880,
  \dodoi{10.1086/321495}

\bibitem[{{Panjkov} {et~al.}(2024){Panjkov}, {Auchettl}, {Shappee}, {Do},
  {Lopez}, \& {Beacom}}]{2024PASA...41...59P}
{Panjkov}, S., {Auchettl}, K., {Shappee}, B.~J., {et~al.} 2024, \pasa, 41,
  e059, \dodoi{10.1017/pasa.2024.66}

\bibitem[{{Payne} \& {Melatos}(2004)}]{2004MNRAS.351..569P}
{Payne}, D.~J.~B., \& {Melatos}, A. 2004, \mnras, 351, 569,
  \dodoi{10.1111/j.1365-2966.2004.07798.x}

\bibitem[{{Pinto} \& {Eastman}(2000)}]{2000ApJ...530..744P}
{Pinto}, P.~A., \& {Eastman}, R.~G. 2000, \apj, 530, 744,
  \dodoi{10.1086/308376}

\bibitem[{{Podsiadlowski} {et~al.}(1992){Podsiadlowski}, {Joss}, \&
  {Hsu}}]{1992ApJ...391..246P}
{Podsiadlowski}, P., {Joss}, P.~C., \& {Hsu}, J.~J.~L. 1992, \apj, 391, 246,
  \dodoi{10.1086/171341}

\bibitem[{Prentice {et~al.}(2016)Prentice, Mazzali, Pian, Gal-Yam, Kulkarni,
  Rubin, Corsi, Fremling, Sollerman, Yaron, Arcavi, Zheng, Kasliwal,
  Filippenko, Cenko, Cao, \& Nugent}]{10.1093/mnras/stw299}
Prentice, S.~J., Mazzali, P.~A., Pian, E., {et~al.} 2016, Monthly Notices of
  the Royal Astronomical Society, 458, 2973, \dodoi{10.1093/mnras/stw299}

\bibitem[{Prentice {et~al.}(2018)Prentice, Ashall, James, Short, Mazzali,
  Bersier, Crowther, Barbarino, Chen, Copperwheat, Darnley, Denneau,
  Elias-Rosa, Fraser, Galbany, Gal-Yam, Harmanen, Howell, Hosseinzadeh,
  Inserra, Kankare, Karamehmetoglu, Lamb, Limongi, Maguire, McCully,
  Olivares E, Piascik, Pignata, Reichart, Rest, Reynolds, Rodríguez, Saario,
  Schulze, Smartt, Smith, Sollerman, Stalder, Sullivan, Taddia, Valenti,
  Vergani, Williams, \& Young}]{10.1093/mnras/sty3399}
Prentice, S.~J., Ashall, C., James, P.~A., {et~al.} 2018, Monthly Notices of
  the Royal Astronomical Society, 485, 1559, \dodoi{10.1093/mnras/sty3399}

\bibitem[{{Rastinejad} {et~al.}(2025){Rastinejad}, {Levan}, {Jonker},
  {Kilpatrick}, {Fryer}, {Sarin}, {Gompertz}, {Liu}, {Eyles-Ferris}, {Fong},
  {Burns}, {Gillanders}, {Mandel}, {Malesani}, {O'Brien}, {Tanvir}, {Ackley},
  {Aryan}, {Bauer}, {Bloemen}, {de Boer}, {Bom}, {Chac{\'o}n}, {Chambers},
  {Chen}, {Chrimes}, {van Dalen}, {D'Elia}, {De Pasquale}, {Fulton}, {Groot},
  {Gupta}, {Hartmann}, {van Hoof}, {Huber}, {Izzo}, {Jacobson-Galan},
  {Jakobsson}, {Kong}, {Laskar}, {Lowe}, {Magnier}, {Maiorano},
  {Martin-Carrillo}, {Mas-Ribas}, {Mata S{\'a}nchez}, {Nicholl}, {Nixon},
  {Oates}, {Paek}, {Palmerio}, {Paris}, {Pieterse}, {Pugliese}, {Quirola
  Vasquez}, {van Roestel}, {Rossi}, {Rouco Escorial}, {Salvaterra},
  {Schneider}, {Smartt}, {Smith}, {Smith}, {Srivastav}, {Torres}, {Ventura},
  {Vreeswijk}, {Wainscoat}, {Yang}, \& {Yang}}]{2025ApJ...988L..13R}
{Rastinejad}, J.~C., {Levan}, A.~J., {Jonker}, P.~G., {et~al.} 2025, \apjl,
  988, L13, \dodoi{10.3847/2041-8213/ade7f9}

\bibitem[{{Reynolds} {et~al.}(2025{\natexlab{a}}){Reynolds}, {Nagao},
  {Gottumukkala}, {Guti{\'e}rrez}, {Kangas}, {Kravtsov}, {Kuncarayakti},
  {Maeda}, {Elias-Rosa}, {Fraser}, {Kotak}, {Mattila}, {Pastorello}, {Pessi},
  {Cai}, {Fynbo}, {Kawabata}, {Lundqvist}, {Matilainen}, {Moran}, {Reguitti},
  {Taguchi}, \& {Yamanaka}}]{2025arXiv250113619R}
{Reynolds}, T.~M., {Nagao}, T., {Gottumukkala}, R., {et~al.}
  2025{\natexlab{a}}, arXiv e-prints, arXiv:2501.13619,
  \dodoi{10.48550/arXiv.2501.13619}

\bibitem[{{Reynolds} {et~al.}(2025{\natexlab{b}}){Reynolds}, {Nagao}, {Maeda},
  {Elias-Rosa}, {Fraser}, {Guti{\'e}rrez}, {Kangas}, {Kuncarayakti}, {Mattila},
  \& {Pessi}}]{2025arXiv250113621R}
{Reynolds}, T.~M., {Nagao}, T., {Maeda}, K., {et~al.} 2025{\natexlab{b}}, arXiv
  e-prints, arXiv:2501.13621, \dodoi{10.48550/arXiv.2501.13621}

\bibitem[{{Sedov}(1959)}]{1959sdmm.book.....S}
{Sedov}, L.~I. 1959, {Similarity and Dimensional Methods in Mechanics}
  (Academic Press)

\bibitem[{{Seitenzahl} {et~al.}(2013){Seitenzahl}, {Ciaraldi-Schoolmann},
  {R{\"o}pke}, {Fink}, {Hillebrandt}, {Kromer}, {Pakmor}, {Ruiter}, {Sim}, \&
  {Taubenberger}}]{2013MNRAS.429.1156S}
{Seitenzahl}, I.~R., {Ciaraldi-Schoolmann}, F., {R{\"o}pke}, F.~K., {et~al.}
  2013, \mnras, 429, 1156, \dodoi{10.1093/mnras/sts402}

\bibitem[{{Smith}(2017)}]{2017hsn..book..403S}
{Smith}, N. 2017, in Handbook of Supernovae, ed. A.~W. {Alsabti} \&
  P.~{Murdin}, 403, \dodoi{10.1007/978-3-319-21846-5_38}

\bibitem[{{Surman} {et~al.}(2006){Surman}, {McLaughlin}, \&
  {Hix}}]{2006ApJ...643.1057S}
{Surman}, R., {McLaughlin}, G.~C., \& {Hix}, W.~R. 2006, \apj, 643, 1057,
  \dodoi{10.1086/501116}

\bibitem[{{Tan} {et~al.}(2001){Tan}, {Matzner}, \&
  {McKee}}]{2001ApJ...551..946T}
{Tan}, J.~C., {Matzner}, C.~D., \& {McKee}, C.~F. 2001, \apj, 551, 946,
  \dodoi{10.1086/320245}

\bibitem[{{Vance} {et~al.}(2020){Vance}, {Young}, {Fryer}, \&
  {Ellinger}}]{2020ApJ...895...82V}
{Vance}, G.~S., {Young}, P.~A., {Fryer}, C.~L., \& {Ellinger}, C.~I. 2020,
  \apj, 895, 82, \dodoi{10.3847/1538-4357/ab8ade}

\bibitem[{{Wheeler} {et~al.}(2000){Wheeler}, {Yi}, {H{\"o}flich}, \&
  {Wang}}]{2000ApJ...537..810W}
{Wheeler}, J.~C., {Yi}, I., {H{\"o}flich}, P., \& {Wang}, L. 2000, \apj, 537,
  810, \dodoi{10.1086/309055}

\bibitem[{{Wollaeger} \& {van Rossum}(2014)}]{2014ApJS..214...28W}
{Wollaeger}, R.~T., \& {van Rossum}, D.~R. 2014, \apjs, 214, 28,
  \dodoi{10.1088/0067-0049/214/2/28}

\bibitem[{{Wollaeger} {et~al.}(2013){Wollaeger}, {van Rossum}, {Graziani},
  {Couch}, {Jordan}, {Lamb}, \& {Moses}}]{2013ApJS..209...36W}
{Wollaeger}, R.~T., {van Rossum}, D.~R., {Graziani}, C., {et~al.} 2013, \apjs,
  209, 36, \dodoi{10.1088/0067-0049/209/2/36}

\bibitem[{{Wollaeger} {et~al.}(2019){Wollaeger}, {Fryer}, {Fontes}, {Lippuner},
  {Vestrand}, {Mumpower}, {Korobkin}, {Hungerford}, \&
  {Even}}]{2019ApJ...880...22W}
{Wollaeger}, R.~T., {Fryer}, C.~L., {Fontes}, C.~J., {et~al.} 2019, \apj, 880,
  22, \dodoi{10.3847/1538-4357/ab25f5}

\bibitem[{{Woosley}(1993)}]{1993ApJ...405..273W}
{Woosley}, S.~E. 1993, \apj, 405, 273, \dodoi{10.1086/172359}

\bibitem[{{Zhang} \& {M{\'e}sz{\'a}ros}(2001)}]{2001ApJ...552L..35Z}
{Zhang}, B., \& {M{\'e}sz{\'a}ros}, P. 2001, \apjl, 552, L35,
  \dodoi{10.1086/320255}

\end{thebibliography}
\bibliographystyle{aasjournal}

\end{document}